\documentclass[a4paper,11pt]{article}
\usepackage{jheppub}
\usepackage{babel}
\usepackage{amsmath,amssymb}
\usepackage{natbib}
\usepackage{graphicx,subfigure}
\usepackage{hyperref}
\usepackage{braket}
\usepackage{comment}
\usepackage[normalem]{ulem}

\newcommand{\be}{\begin{equation}}
 \newcommand{\ee}{\end{equation}}
 \newcommand{\bse}{\begin{subequations}}
 \newcommand{\ese}{\end{subequations}}
\newcommand{\ba}{\begin{eqnarray}}
\newcommand{\ea}{\end{eqnarray}}

\usepackage{color}
\usepackage[normalem]{ulem}  

\begin{document}

\title{Generalized conformal quantum mechanics as an ideal observer in two-dimensional gravity}
\author[a,b,c]{Archi Banerjee,}
\author[d]{Tanay Kibe,}
\author[e]{Mart\'{\i}n Molina,}
\author[e,d]{and Ayan Mukhopadhyay}
\affiliation[a]{Department of Physics, Indian Institute of Technology Kharagpur, Kharagpur 721302, India.}
\affiliation[b]{Max Planck Institute for the Physics of Complex Systems, N{\"o}thnitzer Strasse 38, 01187 Dresden, Germany.}
\affiliation[c]{SUPA, School of Physics and Astronomy, University of St Andrews, St Andrews KY16 9SS, United Kingdom}
\affiliation[d]{Department of Physics, Indian Institute of Technology Madras, Chennai 600036, India.}
\affiliation[e]{Instituto de F\'{\i}sica,
Pontificia Universidad Cat\'{o}lica de Valpara\'{\i}so,
Avenida Universidad 330, Valpara\'{\i}so, Chile.}
\emailAdd{archib@pks.mpg.de, tanaykibe2.71@gmail.com, \\ martinmolinaramos95@gmail.com, ayan.mukhopadhyay@pucv.cl}
\abstract{We obtain an action for a generalized conformal mechanics (GCM) coupled to Jackiw-Teitelboim (JT) gravity from a double scaling limit of the motion of a charged massive particle in the near-horizon geometry of a \textit{near-extremal} spherical black hole. When JT gravity is treated in the classical approximation, the backreaction of the particle's wavefunction on the time-reparametrization mode (and therefore the bulk metric) vanishes while the conformal symmetry in GCM is reparametrized in a state-dependent way. We also construct the semi-classical Hilbert space of the full theory by explicitly solving the general time-dependent normalizable solutions of the Schr\"{o}dinger equation for GCM, and show that the time-reparametrization mode can be inferred from the measurement of suitable observables. Since the full theory of the GCM coupled to JT gravity is amenable to quantization, it can lead to a solvable model for a detector coupled to quantum gravity. }

\maketitle

\section{Introduction}

Jackiw-Teitelboim (JT) gravity \cite{Teitelboim:1983ux,Jackiw:1984je} is one of the simplest gravitational theories which has black hole solutions and admits holographic interpretation \cite{Almheiri:2014cka,Maldacena:2016hyu,Jensen:2016pah,Maldacena:2016upp,Engelsoy:2016xyb,Joshi:2019wgi,Mertens:2022irh}. Furthermore, JT gravity coupled to quantum matter fields gives a simple model in which we can compute the Page curve of an evaporating black hole explicitly \cite{AEMM,AlmheiriQES,Mertens:2022irh}, and it is also a theory which is amenable to full quantization \cite{Penington:2023dql,Kolchmeyer:2023gwa}. Therefore, it is interesting to ask if we can naturally couple a detector to JT gravity for measuring gravitational observables such that the full theory with the detector coupled to JT gravity can be quantized. Such a setup would allow us to \textit{extend the theory of strong and weak measurements} \cite{Jacobs2006,Quanta12} \textit{to quantum gravity.} This would also give deeper insights into \textit{how semi-classical geometries} such as classical black holes with horizons (entropies) \textit{emerge} via self-averaging \textit{from the point of view of a measuring device.}

In this work, we construct a simple model of a particle in a near-extremal throat, and show that the full system can be reduced to generalized conformal mechanics (GCM) coupled to JT gravity in a double scaling limit. This model is amenable to quantization. Here we study the semi-classical limit treating JT gravity classically, and show that the time reparametrization mode (the only degree of freedom in pure JT gravity) can be inferred from suitable observables in the quantized GCM.

Firstly, we examine a charged particle in the near-horizon geometry of a nearly extremal spherical black hole in four dimensions, and show that the action for the probe particle reduces to a (non-relativistic) GCM action under a double scaling limit. Our results generalize the derivation of conformal mechanics (CM) \cite{Case:1950an,deAlfaro:1976vlx} from the double scaling limit of the action of an almost supersymmetric probe particle in an extremal throat \cite{Claus:1998ts,Britto-Pacumio:1999dnb}. This GCM action involves a very specific coupling of the CM to the time-reparametrization mode of JT gravity that describes \cite{Nayak:2018qej,Moitra:2018jqs,Sachdev:2019bjn,Mertens:2022irh} the dynamics of the two-dimensional (nearly) AdS$_2$ factor of the near-horizon geometry. In GCM, the conserved charges are a time-dependent automorphism of the SL(2,R) charges of the usual conformal mechanics. Thus the time-reparametrization mode of the throat determines the \textit{instantaneous SL(2,R) frame} of the (non-relativistic) probe particle much like how gravity generally determines local inertial frames of relativistic particles.\footnote{Time-reparametrization symmetry is broken to SL(2,R) in JT gravity. A solution of JT gravity further breaks SL(2,R) to U(1).} Furthermore, in the full theory in which GCM (describing the particle) is coupled to JT gravity (describing the two-dimensional near-extremal throat), the particle remarkably does not backreact on the gravitational sector in the large N limit (in which the latter can be treated classically).

In order to develop a model of a quantum detector coupled to gravity, as a first step we study the quantization of the GCM coupled to classical JT gravity. The quantization is non-trivial as the GCM has no normalizable ground state as long as the Arnowitt-Deser-Misner (ADM) mass of the classical gravitational solution is positive (containing a dynamical black hole in this case). Nevertheless, we are able to construct general time-dependent normalizable solutions of the Schr\"{o}dinger equation with the boundary condition that the wavefunction vanishes outside the throat following the methods of de Alfaro, Fubini and Furlan \cite{deAlfaro:1976vlx} along with some additional inputs. We demonstrate that the time-reparametrization mode can be inferred from the weak measurements of suitable observables of the quantized GCM theory without affecting the gravitational sector in the large N limit. Therefore, the quantized GCM is an ideal observer/detector for classical JT gravity and motivates the understanding of the full quantum theory where JT gravity should be quantized.\footnote{Here we are not explicit about how a weak measurement of the quantized GCM itself can be done as for this we have standard theories. Such a detector should couple only to the GCM and not to JT gravity. We will thus regard the GCM itself as a detector. We will comment more on this in Sec. \ref{Sec:Detect}.} 

In the concluding section, we discuss that it could be possible to extend our simple theory to a more elaborate setup involving multiple near-extremal throats where the motion of the center of the throats are governed by a generalization of conformal mechanics involving multiple particles. Motivated by the fragmentation instability of the near-extremal near horizon geometries of black holes \cite{Maldacena:1998uz}, we can build tractable microstate models similar to \cite{PhysRevD.102.086008,Kibe:2021gtw,Kibe:2023ixa}, and which could be amenable to quantization. Such setups can help in understanding the microscopic realization of the black hole complementarity principle \cite{PhysRevD.48.3743,PhysRevD.50.2700,Harlow:2014yka,Raju:2020smc,Kibe:2021gtw}.

The plan of the paper is as follows. In Sec. \ref{Sec:GCM}, we discuss the derivation of generalized conformal mechanics from the study of a probe particle in a near extremal throat, and we study the properties of the action. Furthermore, we study the full theory of the generalized conformal mechanics coupled to the JT gravity of the throat. In Sec. \ref{Sec:QGCM}, we study the quantized generalized conformal mechanics and the full semi-classical theory in which JT gravity is treated classically. We also show that the time-reparametrization mode can be inferred from measurements on the generalized conformal mechanics. Finally, we conclude in Sec. \ref{Sec:Disc} with comments on some further extensions of our work which are worthy of study in the future especially from the point of view of obtaining tractable models of quantum black hole microstates.

\section{Generalized conformal mechanics and two-dimensional gravity}\label{Sec:GCM}

\subsection{Derivation from motion of particles in near-extremal throats}\label{Sec:GCM-deriv}
The near horizon geometry of an extremal Reissner-Nordst{\"o}rm spherical black hole in 4 dimensions is AdS$_2 \times S_2$ endowed with an electromagnetic field strength of constant magnitude. The metric and gauge field of the two-dimensional factor take the form:
\begin{equation}\label{Eq:EMSol1}
    {\rm d}s^2 = \frac{4 L^2}{r^2}{\rm d}r^2 -\left(\frac{4 L^2}{r^2}\right)^2{\rm d}t^2, \quad A = \frac{4L^2}{r^2}{\rm d}t.
\end{equation}
The boundary of AdS$_2$ is at $r =0$ while the gauge field vanishes at the Poincar\'{e} horizon which is at $r = \infty$. The Ricci scalar of AdS$_2$ and the electromagnetic field strength are of the same magnitude. Explicitly,
\begin{equation}
    F_{\alpha\beta}F^{\alpha\beta} = R = -\frac{2}{L^2}.
\end{equation}
It is useful to note that the mass $M^{(4)}_{BH}$ of the four dimensional spherical extremal black hole is precisely $L$, the AdS$_2$ radius, in units where the four dimensional Newton's constant $G_N^{(4)} $ is unity.\footnote{A similar discussion can be extended to asymptotically five and higher dimensional spherical extremal black holes -- see \cite{Britto-Pacumio:1999dnb}. More generally, $L\sim {M^{(D)}_{BH}}^{1/(D-3)}$ for $D\geq 4$. In presence of a negative cosmological constant, such a two-dimensional description also arises for the extremal rotating BTZ black hole in three dimensions whose near horizon geometry is AdS$_2$ $\times$ S$^1$ \cite{Ghosh:2019rcj}, etc.}

Consider a particle with mass $M$ and charge $Q$ living in the background \eqref{Eq:EMSol1}. In the static gauge, the Hamiltonian of the particle in this background is $H = - p_t$, where $p_t$ should solve the mass-shell condition
\be\label{Eq:Mass-Shell}
(p_\mu - Q A_\mu)g^{\mu\nu}(p_\nu - Q A_\nu)+ M^2 =0.
\ee
We would be interested to find the Hamiltonian of the particle in the limit in which $L\rightarrow \infty$ and $Q-M\rightarrow 0$ with $L^2(M-Q)$ held fixed. This can be achieved if we parameterize $L$ and $Q-M$ via
\begin{equation}\label{Eq:Limit}
    L = L_0 \epsilon^{-1}, \quad  Q = M - \frac{g}{8L_0^2}\epsilon^2,
\end{equation}
with constant $g$ and take $\epsilon$ to zero. We readily find that the mass shell condition reduces to a linear equation in $p_t$ in this limit, and the Hamiltonian of the particle in the background \eqref{Eq:EMSol1} is \cite{Claus:1998ts}
\begin{equation}\label{Eq:CQMH}
    \mathbb{H}= - p_t = \frac{p_r^2}{2M} + \frac{g}{2 r^2}.
\end{equation}
This is the Hamiltonian of non-relativistic conformal mechanics of a single particle \cite{Case:1950an,deAlfaro:1976vlx}. We end up with a non-relativistic action by virtue of the double scaling limit \eqref{Eq:Limit} which actually amounts to considering a large mass and a large charge of the four dimensional extremal spherical black hole  (since $M^{(4)}_{BH} = L$ in units $G_N^{(4)}=1$). Note also that according to \eqref{Eq:Limit}, the ratio $(M-Q)/M^{(4)}_{BH} $ vanishes as $\epsilon^3$. Since the force on the particle is proportional to $M-Q$, the double scaling limit is a probe limit.  When $M<Q$, $g<0$ and therefore the particle gets repelled by the electric charge of the black hole to $r=0$, the boundary of AdS$_2$. When $M>Q$, the gravitational attraction of the black hole overpowers the electric repulsion and the particle falls into the Poincar\'{e} horizon at $r= \infty$. When $M=Q$, the particle is free since the radial gravitational attraction and radial electric repulsion balance each other out.\footnote{If the particle has angular momentum $l$ in the $S^2$ factor of the four-dimensional near horizon geometry, then $g$ has an additional correction proportional to $l(l+1)$ \cite{Claus:1998ts}. Here we restrict to $l=0$.}

It is useful to find an effective two-dimensional gravitational theory of the near-horizon geometry of the extremal black hole. In fact, the metric and the gauge-field in \eqref{Eq:EMSol1} are solutions of two-dimensional Jackiw-Teitelboim-Maxwell gravity with an additional dilaton field $\phi$. Its action is
\begin{align}\label{Eq:SJTEMD}
    & S =\frac{\phi_0}{16\pi G_N} \int_{\mathcal{M}} \sqrt{-g}\,  R +\frac{\phi_0}{8\pi G_N}\int_{\partial {\mathcal{M}}} \sqrt{-\gamma} K\nonumber\\
    &+\frac{1}{16\pi G_N} \int_{\mathcal{M}} \sqrt{-g}\, \phi\left(R+ \frac{2}{L^2}\right)+\frac{1}{8\pi G_N}\int_{\partial {\mathcal{M}}} \sqrt{-\gamma}\phi_b K\nonumber\\& -\frac{1}{4}\int_{\mathcal{M}} \sqrt{-g}F_{\mu\nu}F^{\mu\nu}.
\end{align}
The terms in the first line above (with $\phi_0$ constant and $G_N$ being the Newton's gravitational constant in two dimensions) is proportional to the Euler characteristic of $\mathcal{M}$.  Above, $\gamma$ is the induced metric at the boundary and $\phi_b$ is the boundary value of the dilaton. The equations of motion are
\begin{align}\label{Eq:JTEom}
     & R + \frac{2}{L^2} =0, \nonumber\\
     & \nabla^\mu F_{\mu\nu} = 0,\nonumber\\
     & \nabla_\mu\nabla_\nu \phi - g_{\mu\nu} \nabla^2 \phi + g_{\mu\nu}\frac{1}{L^2}\phi + 8\pi G_N\left(F_{\mu\lambda}F_{\nu}^{\phantom{\nu}\lambda} - \frac{1}{4}F_{\alpha\beta}F^{\alpha\beta}g_{\mu\nu}\right) =0.
\end{align}
As we will see, the above action has only one effective degree of freedom which is the time-reparametrization mode at the boundary.

It has been shown that the two-dimensional Jackiw-Teitelboim-Maxwell gravity action can indeed be systematically derived from the dimensional reduction of four-dimensional Einstein-Maxwell theory which gives the near-horizon dynamics of a spherical near-extremal four dimensional Reissner-N{\"o}rdstrom black hole \cite{Nayak:2018qej,Moitra:2018jqs,Sachdev:2019bjn,Mertens:2022irh}. The dilaton field in \eqref{Eq:SJTEMD} is essentially the radius of S$^2$ (which is transverse to AdS$_2$). The boundary of AdS$_2$ is effectively where the radius of S$^2$ is an appropriate constant and the Jackiw-Teitelboim gravity description emerges when the fluctuations of the radius of S$^2$ are small compared to its asymptotic value. This effective Jackiw-Teitelboim gravity coupled to gauge fields also captures the entropy of the higher dimensional near extremal black hole \cite{Iliesiu:2020qvm,Heydeman:2020hhw} generalizing the extremal case \cite{Sen:2008vm}.

It is therefore well motivated to study the general solutions of \eqref{Eq:JTEom} which can appear from the dimensional reduction of the near-horizon dynamics of near-extremal higher-dimensional black holes. Furthermore, it is also interesting to find out how the conformal quantum mechanics in \eqref{Eq:CQMH} which appears from the behavior of a  probe particle in the extremal near-horizon geometry generalizes to the near-extremal case. Since the time-reparametrization mode is the only degree of freedom in \eqref{Eq:SJTEMD}, the conformal quantum mechanics in \eqref{Eq:CQMH} should generalize by a suitable coupling of the position of the particle to the time-reparametrization mode. In the rest of this subsection, we derive how the conformal mechanics of the particle is generalized by the coupling to the time reparametrization mode.

To proceed further, we need to find the general solutions of JT-Maxwell gravity. The first equation of motion in \eqref{Eq:JTEom} implies that the metric should be locally AdS$_2$. In order to facilitate comparison with the extremal limit \eqref{Eq:EMSol1}, we choose radial coordinate $\rho$ and time coordinate $u$ such that 
\begin{equation}
    g_{\rho\rho} = \frac{4L^2}{\rho^2}, \quad g_{\rho u} =0.
\end{equation}
Furthermore, the boundary metric should be the same as in \eqref{Eq:EMSol1}. Then the locally AdS$_2$ metric then takes the general form
\begin{eqnarray}\label{Eq:EMSol21}
{\rm d}s^2 &=& \frac{4 L^2}{\rho^2}{\rm d}\rho^2 -\left(\frac{4 L^2}{\rho^2}+\frac{\rho^2}{8}{\rm Sch }(f(u),u)\right)^2{\rm d}u^2.
\end{eqnarray}
where
\begin{equation}
    {\rm Sch }(f(u),u) = \frac{f'''(u)}{f'(u)}- \frac{3}{2}\left(\frac{f''(u)}{f'(u)}\right)^2.
\end{equation}
The metric \eqref{Eq:EMSol21} can be obtained from \eqref{Eq:EMSol1} using the coordinate transformations given by
\begin{align}\label{Eq:CoordTransf}
    & r = \rho \sqrt{f'(u)}\,\,F\left(\dfrac{\rho^4 f''(u)^2}{L^2 f'(u)^2}\right), \quad t = f(u)+ \frac{\rho^4 f''(u)}{32 L^2}\,\, F\left(\dfrac{\rho^4 f''(u)^2}{L^2 f'(u)^2}\right)^2, \nonumber\\
    & F(x) = \left(1-\frac{x}{64}\right)^{-\frac{1}{2}}.
\end{align}
Note that $\rho$ and $u$ coincide with $r$ and $t$ respectively when $f(u) =u$. Furthermore, the SL(2,R) transformation 
\begin{equation}\label{Eq:SL2R}
    f(u) \rightarrow \frac{a f(u) +b}{c f(u) +d}, \quad ad - bc =1
\end{equation}
keeps ${\rm Sch}(f(u),u)$ invariant and therefore the metric \eqref{Eq:EMSol21} also. Therefore, the different $\rho$ and $u$ obtained from \eqref{Eq:CoordTransf} and \eqref{Eq:SL2R} give isometries of the metric \eqref{Eq:EMSol22}. We readily note from \eqref{Eq:CoordTransf} that at the boundary $\rho = 0$, the time-coordinates of \eqref{Eq:EMSol1} and \eqref{Eq:EMSol21} are related by $t = f(u)$. We will explicitly see from the solution of the dilaton later in this subsection and the renormalized on-shell action of the gravitational theory in Sec. \ref{Sec:FT} that $f(u)$, the time reparametrization mode, is a degree of freedom of the theory (modulo SL(2,R) transformations).

We also note that the Poincar\'{e} horizon $r= \infty$ maps to
\begin{equation}\label{Eq:Horrho}
    \rho = \rho_h(u) = 2\sqrt{2 L}\sqrt{\frac{f'(u)}{f''(u)}}
\end{equation}
where $F$ diverges and indeed $\rho = \rho_h(u)$ is a Killing horizon.  For a concrete example, we can choose 
\begin{equation}\label{Eq:fBH}
    f(u) = L \exp\left(\frac{2\pi u}{\beta}\right),
\end{equation}
for which
\begin{equation}
    {\rm Sch}(f(u),u) = - \frac{2\pi^2}{\beta^2}.
\end{equation}
The solution \eqref{Eq:EMSol21} then corresponds to a two-dimensional static black hole with mass
\begin{equation}
    M_{\rm BH} \sim -2\, {\rm Sch}(f(u),u) = \frac{4\pi^2}{\beta^2}
\end{equation}
up to a factor of $N^2\sim G_N^{-1}$ and Hawking temperature $\beta^{-1}$ (note from \eqref{Eq:fBH} that the periodicity of the Euclidean time circle is indeed $\beta^{-1})$. Furthermore $\rho_h(u)$ in \eqref{Eq:Horrho} then corresponds to the Killing horizon where $g_{uu}$ vanishes. In this case. explicitly
\begin{equation}\label{Eq:BHhor}
    \rho_h(u) = 2 \sqrt{\frac{L\beta}{\pi}}.
\end{equation}

We examine the general solutions of the gauge field. In order to compare with \eqref{Eq:EMSol1}, we fix the $U(1)$ gauge by requiring that
\begin{equation}
    A_\rho =0.
\end{equation}
Furthermore, requiring that $A_u$ vanishes at the Killing horizon $\rho = \rho_h(u)$ (with  $\rho_h(u)$ given by \eqref{Eq:Horrho})\footnote{In the case of the static black hole \eqref{Eq:fBH}, this condition implies the contractibility of the Wilson loop on the time circle after Euclidean rotation.}, the general solution for the gauge field turns out to be
\begin{equation}\label{Eq:EMSol22}
    A =\frac{K}{32 L^2}\left(\rho_h(u)^2 - \rho^2\right)\left({\rm Sch}(f(u),u)+ \frac{4 L}{\rho^2}\frac{f''(u)}{f'(u)}\right){\rm d}u.
\end{equation}
The magnitude of the electromagnetic field strength is
\begin{equation}\label{Eq:K4Lsq}
    F_{\alpha\beta}F^{\alpha\beta} = -\frac{K^2}{8 L^6}, \quad {\rm and}\quad K- 4 L^2 = 2\sqrt{2}L^3 \left(\sqrt{-F_{\alpha\beta}F^{\alpha\beta}} - \sqrt{-R} \right).
\end{equation}
Note that the above reduces to the gauge field in \eqref{Eq:EMSol1} when $K=4L^2$ and $f(u) =u$ (so that $\rho$ and $u$ also coincide with $r$ and $t$ respectively following \eqref{Eq:CoordTransf}). Since, the two-dimensional solution is expected to be obtained as a near-horizon geometry of a non-extremal higher dimensional black hole, we do not need to set $K = 4L^2$ as in the extremal limit. When $K = 4 L^2$, we can obtain \eqref{Eq:EMSol22} from \eqref{Eq:EMSol1} after performing a further $U(1)$ gauge transformation 
\begin{equation}
   A\rightarrow A + d\lambda,\quad \lambda(\rho,u) =L \ln\left(\dfrac{1 - \frac{\rho^2 f''(u)}{8Lf'(u)}}{1 + \frac{\rho^2 f''(u)}{8Lf'(u)}}\right)+\frac{K}{4L}\ln\left(\frac{L f''(u)}{f'(u)^2}\right)
\end{equation}
following the coordinate transformations \eqref{Eq:CoordTransf} so that we retain the gauge $A_\rho =0$ and the boundary condition that $A_u$ vanishes at the Killing horizon.

Furthermore, we note that the gauge field configuration \eqref{Eq:EMSol22} preserves SL(2,R) symmetry. Recall that a SL(2,R) transformation of $f(u)$ corresponds to an isometry of the background metric. We can verify that the transformation of \eqref{Eq:EMSol22} under an SL(2,R) isometry can be compensated by a $U(1)$ gauge transformation so that we retain the form \eqref{Eq:EMSol22} for the new $f(u)$ in which $A_\rho = 0$ and $A_u$ vanishes at the Killing horizon.

Finally, the general solution for the dilation equation of motion (the last equation in \eqref{Eq:JTEom}) is given by
\begin{align}\label{phi-sol}
 & \phi(\rho,u) = \frac{\alpha_1(u)}{\rho^2}+\frac{K^2}{32 L^4}+ \rho^2 \alpha_2(u),\nonumber\\
 & \alpha_1(u) = \frac{a_1 + a_2 f(u) + a_3 f(u)^2}{f'(u)},\nonumber\\
 & \alpha_2(u) = - \frac{a_3}{16L^2} f'(u) + \frac{1}{32 L^2}\frac{f''(u)}{f'(u)}\left(a_2 + 2 a_3 f(u)\right)\nonumber\\& \qquad\qquad-\frac{1}{64 L^2}\frac{f''(u)^2}{f'(u)^3}\left(a_1 + a_2 f(u)+ a_3 f(u)^2\right).
\end{align}
Above, $a_1$, $a_2$ and $a_3$ are arbitrary constants. For a given choice of these constants, we note that the dilaton field is determined by $f(u)$. The latter can be obtained by setting a boundary value of the dilation $\phi_b(u)$ on a cut-off hypersurface. We will discuss this more explicitly in Sec. \ref{Sec:FT}.

We recall that a SL(2.R) transformation of $f(u)$ corresponds to an isometry of the metric \eqref{Eq:EMSol21} and therefore preserves the form of the dilaton solution \eqref{phi-sol}. Therefore, any SL(2,R) transformation of $f(u)$ corresponds only to redefinitions of $a_1$, $a_2$ and $a_3$. Nevertheless, the dilaton configuration \eqref{phi-sol} does not break the SL(2,R) symmetry fully. It can be readily checked that SL(2.R) transformations of $f(u)$ do not change $a_1/a_3$. Since there are three parameters in a generic SL(2,R) transformation and there are only two independent combinations of $a_1$, $a_2$ and $a_3$ which can vary, it follows that there should be a one parameter family of SL(2,R) transformations which does not affect the dilaton profile \eqref{phi-sol}. This implies that a $U(1)$ symmetry is retained. One can also see this from the dilaton equation of motion in \eqref{Eq:JTEom}. Since $F^2$ is a constant, it follows that $\nabla^2\phi$ is a constant and $\nabla_\mu\nabla_\nu \phi$ is a constant times the metric. Therefore, $\epsilon^{\mu\nu}\nabla_\nu \phi$ is a Killing vector implying the presence of a $U(1)$ symmetry. Since the gauge field configuration \eqref{Eq:EMSol22} does not break SL(2,R) symmetry as noted before, the full solution retains a $U(1)$ symmetry.  A version of this argument has been given for pure JT gravity in \cite{Maldacena:2016upp}.

Let us consider a particle with mass $M$ and charge $Q$ in the background \eqref{Eq:EMSol21} and \eqref{Eq:EMSol22}. Generalizing \eqref{Eq:Limit} which has been previously done in the extremal limit, we parametrize
\begin{align}\label{Eq:Limit2}
    L = \frac{L_0}{\epsilon^2},\quad K = 4\frac{L_0^2}{\epsilon^4} +k,\quad Q = M - \frac{1}{8L_0^2}(g + 2 Mk)\epsilon^2
\end{align}
and take the limit $\epsilon \rightarrow 0$. Since $K - 4L^2$ is finite when $\epsilon \rightarrow 0$ (and is the constant $k$), the scaling of $K$ above can be understood as taking a near-extremal limit. Note from \eqref{Eq:K4Lsq} that $K - 4L^2$ is proportional to the difference between the square root of the magnitude of the electromagnetic field strength and the square root of the magnitude of the Ricci scalar which equalizes only in the extremal limit.

In the above limit, the mass shell condition \eqref{Eq:Mass-Shell} in the background metric \eqref{Eq:EMSol21} and gauge field \eqref{Eq:EMSol22} reduces to a linear equation for $p_t$ which can be readily solved to obtain the Hamiltonian 
\begin{equation}
    \mathbb{H} =-p_t= \frac{p_\rho^2}{2M} + \frac{g}{2 \rho^2}+ \frac{\rho^2}{4}M\,{\rm Sch }(f(u),u) - M\xi'(u),
\end{equation}
with 
\begin{equation}
    \xi(u) = \frac{K}{4L}\ln\left(\frac{L f''(u)}{f'(u)^2}\right).
\end{equation}
It is useful to do the canonical transformations
\begin{equation}
    x = \rho\sqrt{M}, \quad p = \frac{p_\rho}{\sqrt{M}}.
\end{equation}
We then obtain
\begin{equation}\label{Eq:HamilN}
    \mathbb{H} = \frac{p^2}{2} + \frac{q}{2 x^2}+ \frac{x^2}{4}\,{\rm Sch }(f(u),u)- M\xi'(u),
\end{equation}
where
\begin{equation}\label{Eq:qNCQM}
    q = gM = \lim_{\epsilon\rightarrow 0}M\left(8 L^2 (M- Q) - (K - 4 L^2)(Q+M)\right).
\end{equation}
In units $\hbar =1$, $q$ is dimensionless, $x$ has mass dimension $-1/2$ and $p$ has mass dimension $1/2$.

We note that unlike the extremal case, the sign of $q$ is not only determined by whether $M$ is greater than or lesser than $Q$, but also by whether $K$ is greater than or lesser than $4L^2$, i.e. whether the background electromagnetic field strength is greater than or lesser than the background Ricci scalar. When $f(u) =u$ (up to a SL(2,R) transformation), ${\rm Sch }(f(u),u)$ vanishes and the particle is free when
\begin{equation}
    \frac{M-Q}{M+Q} = \frac{K-4L^2}{8L^2}
\end{equation}
as $q$ vanishes. We recall that when $f(u)$ is given by \eqref{Eq:fBH}, ${\rm Sch }(f(u),u) <0$ and constant, and the metric corresponds to a static two-dimensional black hole. We note from \eqref{Eq:HamilN}, that the potential is negative and divergent at $x=\infty$ (i.e. $\rho =\infty$) and it creates a bottomless ``well'' where the particle can be potentially sucked into. One can similarly discuss other cases. 

We will see in Sec. \ref{Sec:QGCM} that the quantum theory makes sense even when ${\rm Sch }(f(u),u) <0$. This is because following the approach of \cite{deAlfaro:1976vlx} for the usual conformal mechanics, it can be shown that there exists a conserved charge. Evolving with this conserved charge gives time-dependent normalizable solutions of the Schr{\"o}dinger equation with the original Hamiltonian \eqref{Eq:HamilN}. The definition of this conserved charge involves a length scale which acts as an infrared regulator ensuring that the wavefunction always falls off at large $x$, i.e. large $\rho$ and also vanishes at the boundary $x=0$ (i.e. $\rho =0$). From the bulk point of view, evolving via this conserved charge is evolving along  orbits of a time-like Killing vector which does not become null in the domain $0 < \rho < \infty$ where the wavefunction is supported, generalizing the same result \cite{Claus:1998ts} for usual conformal mechanics.

\subsection{Action, classical solutions and conserved charges}\label{Sec:GCM-action}
\label{Sec:actionandsoln}
The Hamiltonian \eqref{Eq:HamilN} corresponds to the action
\begin{equation}\label{Eq:SNCQM}
    S_{GCM} = \int {\rm d}u \left(\frac{1}{2}x'(u)^2 - \frac{q}{2 x(u)^2}- \frac{1}{4}x(u)^2\,{\rm Sch }(f(u),u) \right)
\end{equation}
after discarding the total derivative term. Our derivation shows that the above action for generalized conformal mechanics captures how the time-reparametrization mode of the gravitational background affects the motion of the charged massive (almost supersymmetric) particle in the near-extremal throat. 

This action has SL(2,R) symmetries which generalize those of the usual conformal mechanics. To see this consider the general time reparametrizations: $u\rightarrow t(u)$ with $t'(u)\geq 0$ under which $x(u)$ and $f(u)$ transform as
\begin{align}\label{Eq:Transform}
x(u) \rightarrow {\tilde{x}(t(u))} =\sqrt{t'(u)} \, x(u), \quad f(u) \rightarrow f(t(u)).
\end{align}
We readily note that
\begin{eqnarray}
\frac{{\rm d}t}{\tilde{x}(t(u))^2} &=& \frac{{\rm d}u}{x(u)^2},\label{Eq:Id1}\\
{\rm d}t\,\tilde{x}'(t(u))^2 &=& {\rm d}u\,{x'(u)^2} - \frac{1}{2} {\rm d}u \,\,{\rm Sch}(t(u), u)\,  x(u)^2\nonumber\\ && -{\rm d}u \frac{{\rm d}}{{\rm d}u}\left(\frac{{\rm d}}{{\rm d}u}\left(\frac{1}{\sqrt{t'(u)}}\right)\frac{\tilde{x}(t(u))^2}{\sqrt{t'(u)}}\right)\label{Eq:Id2}.
\end{eqnarray}
The \textit{chain rule} for the Schwarzian derivative is
\begin{equation}\label{Eq:Id34}
 {\rm Sch}(f(t(u)),t(u)) = \frac{1}{t'(u)^2} \left({\rm Sch}(f(t(u)), u) - {\rm Sch}(t(u),u)\right).
\end{equation}
The above identities imply that the Lagrangian in \eqref{Eq:SNCQM} is invariant up to a total derivative provided
\begin{equation}
    {\rm Sch}(f(t(u)), u) = {\rm Sch}(f(u), u),
\end{equation}
i.e. if
\begin{equation}\label{Eq:Transform2}
f(t(u)) = (G(a,b,c) \circ f) (u) \equiv \frac{a f(u) + b}{c f(u) + d}
\end{equation}
where $a$, $b$, $c$ and $d$ are real numbers satisfying $ad -bc =1$ and labelling $G$ which is an element of the SL(2,R) group. It follows that the action \eqref{Eq:SNCQM} is invariant under the SL(2,R) transformations $u\rightarrow t(u)$ with
\begin{equation}\label{Eq:t-u-1}
    t(u) = (f^{-1}\circ G(a,b,c) \circ f)(u) \equiv f^{-1}\left(\frac{a f(u) + b}{c f(u) + d}\right),
\end{equation}
and under which $x(u)$ and $f(u)$ transform according to \eqref{Eq:Transform}. 

The identities \eqref{Eq:Id1}, \eqref{Eq:Id2} and \eqref{Eq:Id34} also imply that action \eqref{Eq:SNCQM} can be obtained from the usual conformal mechanics action given by
\begin{equation}\label{Eq:SCQM}
    S_{CQM} = \int {\rm d}u \left(\frac{1}{2}z'(u)^2 - \frac{q}{2 z(u)^2} \right)
\end{equation}
via the combined field redefinition $$x(u) \rightarrow z(f(u))= x(u)\sqrt{f'(u)}$$ and time reparametrization $$u\rightarrow f(u).$$We can also derive the SL(2,R) symmetry of the action \eqref{Eq:SNCQM} presented above as a time-reparametrized version of the SL(2,R) symmetry of the usual action \eqref{Eq:SCQM} via the map between the two. In fact, the map between $x(u)$ and $z(u)$ is just the near-boundary truncation of the map between the radial coordinate of the extremal and near-extremal throats given by \eqref{Eq:CoordTransf} (recall that $x$ and $z$ are essentially the radial coordinates $\rho$ and $r$ respectively), and similarly $u$ and $f(u)$ are also the boundary time-coordinates of the extremal and near-extremal throats as evident from \eqref{Eq:CoordTransf}. Although the action and the classical equations of motion derived from the actions \eqref{Eq:SCQM} and \eqref{Eq:SNCQM} can be mapped to each other, the quantum theories described these actions are very different.\footnote{Even classically, the generalized conformal mechanics cannot be viewed as usual conformal mechanics unless we let the Hamiltonian generate flow along the reparametrized time $f(u)$ (aside from the field redefinition).} In particular, there is no simple map between the normalizable solutions of the Schr{\"o}dinger equations of the two quantum theories, and they also have distinct properties as should be clear from the discussions in Sec. \ref{Sec:QGCM}.

The equation of motion for $x(u)$ which follows from the action \eqref{Eq:SNCQM} is
\begin{equation}\label{Eq:x-Eom1}
    x''(u)  = \frac{q}{x(u)^3} - \frac{1}{2}{\rm Sch}(f(u),u) \, x(u).
\end{equation}
For later purposes it will be useful to varying the action \eqref{Eq:SNCQM} w.r.t. $f(u)$. Explicitly, this variation gives
\begin{equation}\label{Eq:f-Eom1}
    x(u)^2\frac{{\rm d}}{{\rm d}u}{\rm Sch}(f(u), u)+4 {x'(u)}{x(u)}{\rm Sch}(f(u), u)+ 6 x'(u) x''(u) + 2x'''(u)x(u) = 0.
\end{equation}
Remarkably, this vanishes identically when $x(u)$ is on-shell as can be checked by substituting $x'''(u)$ above by what we obtain from the time-derivative of \eqref{Eq:x-Eom1}. In other words, the action is extremized for an arbitrary $f(u)$ when $x(u)$ is on-shell. This is not surprising as the action \eqref{Eq:SNCQM} is related to the action \eqref{Eq:SCQM} as discussed above via field redefinition and time-reparametrization, and thus $f(u)$ can be absorbed into $z(u)$ which is the only degree of freedom in \eqref{Eq:SCQM}. This feature will have an important consequence for the full theory of the particle coupled to JT gravity as discussed in the following section.

In order to obtain the solutions of \eqref{Eq:x-Eom1}, we first note that the equation of motion of $z(u)$ obtained from the action \eqref{Eq:SCQM} is
\begin{equation}\label{Eq:x-Eom2}
    z''(u)  = \frac{q}{z(u)^3},
\end{equation}
and its general solutions are \cite{deAlfaro:1976vlx}
\begin{equation}
    z(u) = \sqrt{\frac{q + c_1^2 (u + c_2)^2}{c_1}}.
\end{equation}
The map between $x(u)$ and $z(u)$ discussed above then implies that the general solutions of \eqref{Eq:x-Eom1}  are
\begin{equation}\label{Eq:xSol1}
    x(u) = \frac{z(f(u))}{\sqrt{f'(u)}} =\sqrt{\frac{q + c_1^2 (f(u) + c_2)^2}{c_1 f'(u)}}=\sqrt{\frac{(q+ c_1^2c_2^2)+ 2 c_1^2 c_2 f(u)+c_1^2 f(u)^2}{c_1 f'(u)}}.
\end{equation}
The constants $c_1$ and $c_2$ are determined by initial conditions. Here we will take the view that since $x(u)$ and $z(u)$ are the radial coordinates of the extremal and non-extremal throats (up to proportionality constants), they are both non-negative. At this point we note that $x^2(u)$ takes the same form as the non-normalizable mode ($\alpha_1(u)$) of the on-shell bulk dilaton (see \eqref{phi-sol}) with 
\begin{equation}
     a_1 = \frac{q+ c_1^2c_2^2}{c_1}, \quad a_2 = 2 c_1 c_2, \quad a_3 = c_1,\quad a_1 a_3 - \frac{a_2^2}{4} = q.
\end{equation}
This will be useful for later purposes also.

We readily see that the Hamiltonian \eqref{Eq:HamilN} generates the equation of motion \eqref{Eq:x-Eom1} and
\begin{equation}\label{Eq:HfWI}
    \frac{{\rm d}\mathbb{H}}{{\rm d}u} = \frac{\partial \mathbb{H}}{{\partial}u} = \frac{1}{4}\frac{{\rm d}{\rm Sch}(f(u),u)}{{\rm d}u}x^2(u)
\end{equation}
when $x(u)$ is on-shell, i.e. of the form \eqref{Eq:xSol1}.

Although the Hamiltonian \eqref{Eq:HamilN} is explicitly time-dependent, the SL(2,R) symmetries of the action \eqref{Eq:SNCQM} imply the existence of conserved charges. We proceed to list these conserved charges directly in the quantum theory.

It is useful to first consider these operators
\begin{eqnarray}\label{Eq:h0d0k0}
H_0 = \frac{1}{2}\left(p^2 + \frac{q}{x^2}\right), \quad D_0 = \frac{1}{2}(xp + px), \quad K_0= \frac{1}{2}x^2,
\end{eqnarray}
which form the SL(2,R) Lie algebra
\begin{eqnarray}\label{Eq:Alg1}
[D_0,H_0] = 2 iH_0,\quad [D_0, K_0] =- 2 iK_0, \quad [H_0 ,K_0] = - i D_0
\end{eqnarray}
as can be verified using $[x,p]= i$. 

For any smooth function $f(u)$, there exists an automorphism of the above Lie algebra
\begin{eqnarray}\label{Eq:H1-D1-K1}
\tilde{H}_f &=&  \frac{1}{f'(u)} H_0+\frac{f''(u)}{2{f'(u)}^2} D_0 + \frac{{f''(u)}^2}{4 {f'(u)}^3} K_0,\nonumber\\
\tilde{D}_f &=& D_0 + \frac{f''(u)}{f'(u)}K_0,\nonumber\\
\tilde{K}_f &=& f'(u) K_0,
\end{eqnarray}
so that 
\begin{eqnarray}
[\tilde{D}_f, \tilde{K}_f] =- 2 i\tilde{K}_f, \quad  [\tilde{D}_f,\tilde{H_f}] = 2 i\tilde{H}_f,\quad [\tilde{H}_f ,\tilde{K}_f] = - i \tilde{D}_f.
\end{eqnarray}
Note that $\tilde{H}_f$, $\tilde{D}_f$ and $\tilde{K}_f$ reduce to $H_0$, $D_0$ and $K_0$, respectively, when $f(u) =u$. The above can be verified using \eqref{Eq:h0d0k0}, \eqref{Eq:H1-D1-K1} and $[x,p]= i$.

The conserved Noether charges of the action \eqref{Eq:SNCQM} corresponding to the SL(2,R) symmetries given by \eqref{Eq:Transform} and \eqref{Eq:t-u-1} turn out to be
\begin{eqnarray}\label{Eq:Alg2}
H_f &=& \tilde{H}_f,\nonumber\\
D_f &=&\tilde{D}_f  -2 f(u) \tilde{H}_f ,\nonumber\\
K_f &=& \tilde{K}_f- f(u) \tilde{D}_f+ f(u)^2 \tilde{H}_f .
\end{eqnarray}
The above constants of motion reduce to the conserved charges of usual conformal quantum mechanics (corresponding to the action \eqref{Eq:SCQM}) when we set $f(u) =u$, and furthermore satisfy the same Lie algebra as \eqref{Eq:Alg1} since
\begin{eqnarray}
[D_f, K_f] =- 2 iK_f,\quad [D_f,H_f] = 2 iH_f, \quad [H_f ,K_f] = - i D_f.
\end{eqnarray}
The above can be verified using \eqref{Eq:h0d0k0}, \eqref{Eq:H1-D1-K1}, \eqref{Eq:Alg2} and $[x,p]= i$. As mentioned in the Introduction, one can view generalized conformal mechanics \eqref{Eq:SNCQM} as a \textit{covariantization} of usual conformal mechanics. Quite like how gravity determines the inertial frame of a relativistic particle, here the time reparametrization mode $f(u)$ determines the \textit{instantaneous SL(2,R) frame}, i.e. the conserved charges which furnish the sl(2,R) Lie algebra.

For the general \textit{classical} solutions of the action \eqref{Eq:xSol1}, the conserved charges take the following values in the classical theory: 
 \begin{equation}\label{Eq:OSCC1}
     H_f = \frac{c_1}{2}, \quad D_f = c_1 c_2, \quad K_f = \frac{q}{2 c_1} + \frac{1}{2}c_1 c_2^2.
 \end{equation}
It is important to note that $H_f$ is different from the Hamiltonian of the theory \eqref{Eq:HamilN} unless ${\rm Sch}(f(u),u) =0$ in which case the action coincides with that of usual conformal mechanics.

It is easy to check using \eqref{Eq:h0d0k0}, \eqref{Eq:H1-D1-K1}, \eqref{Eq:Alg2} and $[x,p]=i$ that the Casimir of the sL(2,R) Lie algebra (for arbitrary $f(u)$) is 
\begin{align}\label{Eq:Cas}
   &\mathcal{C}= H_0K_0 - \frac{1}{4}{D_0}^2 = \tilde{H}_f \tilde{K}_f - \frac{1}{4} \tilde{D}_f^2= H_f K_f - \frac{1}{4} D_f^2 
   = \frac{q}{4} -\frac{3}{16}.
\end{align}
We readily note from \eqref{Eq:OSCC1} that $\mathcal{C} = q/4$ on-shell in the classical theory.

For the sake of completeness, we note that the action \eqref{Eq:SNCQM} has another SL(2,R) symmetry under which $f(u)$ transforms as \eqref{Eq:SL2R} while $x(u)$ remains invariant. It turns out that on-shell, the corresponding conserved charges are same in magnitude and opposite to those given by \eqref{Eq:Alg2}. Since the action is extremized for any $f(u)$ when $x(u)$ is on-shell, we do not expect a new set of SL(2,R) conserved charges to exist which generate transformation of $f(u)$ alone (as otherwise it can imply some constraints on $x(u)$ which should be satisfied in addition to its equation of motion). We will see another perspective on this in the next subsection.

\subsection{A nice form for the Hamiltonian of generalized conformal mechanics}\label{Se:Hnice}

The Hamiltonian of the generalized conformal mechanics can be expressed in a nice form. To see this, we first consider the Schwarzian theory
\begin{equation}
    S = \int {\rm d}u \, {\rm Sch}(f(u),u)
\end{equation}
which is invariant under the fractional linear transformations \eqref{Eq:SL2R} of $f(u)$ which give a representation of the SL(2,R) group. The conserved charges of this action are
\begin{align}\label{Eq:Q}
    & Q_0(u) = \frac{f'''(u)}{f'(u)^2} - \frac{f''(u)^2}{f'(u)^3}, \nonumber\\
    & Q_1(u) = -2f(u) \left(\frac{f'''(u)}{f'(u)^2} - \frac{f''(u)^2}{f'(u)^3}\right) +2 \frac{f''(u)}{f'(u)},\nonumber\\
    & Q_2(u) = f(u)^2 \left(\frac{f'''(u)}{f'(u)^2} - \frac{f''(u)^2}{f'(u)^3}\right) -2f(u)\left( \frac{f''(u)}{f'(u)}-\frac{f'(u)}{f(u)}\right).
\end{align}
with $Q_i$ obtained from Noether's procedure considering the infinitesimal versions of the the fractional linear transformations \eqref{Eq:SL2R}:
\begin{equation}\label{Eq:v}
    f(u) \rightarrow f(u) + \epsilon v_i, \quad  v_i = \{1, - 2 f(u), f(u)^2 \},
\end{equation}
respectively. These Noether charges satisfy the Ward identities:
\begin{equation}\label{Eq:QWI}
    \frac{{\rm d}}{{\rm d}u} Q_i = \frac{v_i}{f'(u)} \frac{{\rm d}}{{\rm d}u} {\rm Sch}(f(u),u).
\end{equation}

Under SL(2,R) transformations \eqref{Eq:SL2R}, the vector $$\mathcal{Q}=\{Q_0, Q_1, Q_2\}$$transforms exactly like the conserved charge vector $$\mathcal{G}=\{H_f, D_f, K_f\}$$of generalized conformal mechanics under the transformations given by \eqref{Eq:Transform} and \eqref{Eq:t-u-1} (i.e. in the adjoint representation of SL(2,R)). If $\mathcal{A}$ and  $\mathcal{B}$ are two such adjoint vectors, then we can form a SL(2,R) invariant product
\begin{equation}
    \mathcal{A} \bullet \mathcal{B} =\frac{1}{2} \left(\mathcal{A}_0\mathcal{B}_2+ \mathcal{A}_2\mathcal{B}_0\right) - \frac{1}{4}\mathcal{A}_1 \mathcal{B}_1.
\end{equation}
We readily find that
\begin{equation}\label{Eq:Q2}
     \mathcal{G}\bullet\mathcal{G} = \mathcal{C},\quad \mathcal{Q}\bullet\mathcal{Q} = Q_0 Q_2  - \frac{1}{4}Q_1^2 = 2 {\rm Sch}(f(u),u).
\end{equation}
Above $\mathcal{C}$ is the Casimir defined in \eqref{Eq:Cas}. Note also $v\bullet v =0$.

After some algebra, it is possible to verify using \eqref{Eq:h0d0k0}, \eqref{Eq:H1-D1-K1} and \eqref{Eq:Alg2} that the Hamiltonian of the generalized conformal mechanics \eqref{Eq:HamilN} (without the total derivative term proportional to identity) takes the remarkably simple form
\begin{equation}\label{Eq:Hniceform}
    \mathbb{H} = \mathcal{Q}\bullet\mathcal{G}.
\end{equation}
It is to be noted that $\mathbb{H}$ is not SL(2,R) invariant under the transformations of $\mathcal{G}$ under \eqref{Eq:Transform} and \eqref{Eq:t-u-1} only. It is necessary also to transform $\mathcal{Q}$ under \eqref{Eq:SL2R} simultaneously for the SL(2,R) invariance of the dot product. As should be clear from the discussion in Sec. \ref{Sec:GCM-deriv}, the latter transformations realize the isometries of the two dimensional gravitational metric \eqref{Eq:EMSol21}. Therefore, \eqref{Eq:Hniceform} should be regarded as an invariant of the full theory combining the particle and JT gravity, which will be discussed in detail in the following subsection.\footnote{As shown in the previous subsection, the GCM action is separately invariant under the transformations given by \eqref{Eq:Transform} and \eqref{Eq:t-u-1}, and transformations given by \eqref{Eq:SL2R} where $f(u)$ alone transforms. Furthermore, the Noether charges under these separate SL(2,R) transformations are equal in magnitude while opposite in sign on-shell. So if both the transformations are done simultaneously, as necessary for the SL(2,R) invariance of the Hamiltonian, then the total Noether charges vanish on-shell.}

Another identity, which can also be verified using \eqref{Eq:h0d0k0}, \eqref{Eq:H1-D1-K1} and \eqref{Eq:Alg2}, is
\begin{equation}\label{Eq:x2nice}
    x^2 = \frac{4}{f'(u)} v \bullet \mathcal{G}
\end{equation}
where $v$ is a vector defined in \eqref{Eq:v}.  Given that $\mathcal{G}$ is conserved and that its SL(2,R) norm is the Casimir given by \eqref{Eq:Cas}, we can readily use the above to derive the general solution \eqref{Eq:xSol1} for generalized conformal mechanics.

We can also check that
\begin{equation}
    \frac{{\rm d}\mathbb{H}}{{\rm d}u} = \frac{{\rm d}\mathcal{Q}}{{\rm d}u} \bullet \mathcal{G} = \frac{1}{f'(u)}\left(v \bullet \mathcal{G}\right) \frac{{\rm d}}{{\rm d}u} {\rm Sch}(f(u),u) = \frac{1}{4} x^2 \frac{{\rm d}}{{\rm d}u} {\rm Sch}(f(u),u).
\end{equation}
Above we have used that $\mathcal{G}$ is a conserved charge vector, and also the identities \eqref{Eq:QWI} and \eqref{Eq:x2nice}. Of course, we have just reproduced the Ward identity \eqref{Eq:HfWI}. 

\subsection{The full theory}\label{Sec:FT}

The action for the full theory of the particle coupled to two-dimensional gravity is simply
\begin{equation}
    S_{\rm T} = S_{{\rm GCM}}[x(u), f(u)] + S_{{\rm JT-gen}}^{\rm on-shell}[f(u),J(u)]
\end{equation}
where $S_{{\rm GCM}}$ is the action for generalized conformal mechanics of the particle which depends on $x(u)$ and the time reparametrization mode $f(u)$ of the gravitational sector, and $S_{{\rm JT-gen}}^{\rm on-shell}$ is the regularized action for two-dimensional JT gravitational theory coupled to arbitrary bulk matter where all bulk fields have been expressed in terms of the time reparametrization mode $f(u)$ and $J(u)$, which denotes the sources of bulk matter fields collectively. As shown below, $S_{{\rm JT-gen}}^{\rm on-shell}$ is a boundary action. Note that $f(u)$ appearing in $S_{{\rm GCM}}$ should be identified with the time reparametrization mode of JT gravity, and this is obvious from the derivation of $S_{{\rm GCM}}$ presented in Sec. \ref{Sec:GCM-deriv}.  There is, however, an elegant reformulation of the action using auxiliary fields, where $f(u)$ appearing in $S_{{\rm GCM}}$ and the time reparametrization mode in $S_{{\rm JT-gen}}^{\rm on-shell}$ get identified with each other automatically. This reformulation has been presented in Appendix \ref{Sec:Full2}.

It is useful to also have a holographic perspective for the full action and rewrite it in the form
\begin{equation}\label{Eq:ST-hol}
    S_{\rm T} = S_{{\rm GCM}}  + W_{\rm QT}\left[\phi_r(u) = \phi_0\right] 
\end{equation}
Here, $W_{\rm QT}$ is the logarithm of the partition function (quantum effective action) of a putative large N quantum dot theory dual to a two dimensional gravity theory, where $N^2\sim G_N^{-1}$. Above, $\phi_0$ is an arbitrary constant and $\phi_r(u)$, as explained below, is the source of the operator dual to the bulk dilaton. In the large N limit,
\begin{equation}
    W_{\rm QT} = S_{\rm JT-gen}^{\rm on-shell}
\end{equation}
with $S_{\rm JT-gen}^{\rm on-shell}$ denoting the on-shell gravitational action of JT gravity coupled to matter, where the metric and other bulk fields are determined in terms of the time reparametrization mode via bulk equations of motion. As shown below, this action reproduces the constraint of the gravitational theory amounting to the equation of motion of the time-reparametrization mode (and thus, as expected, encapsulates the full gravitational theory). Generally, $S_{\rm JT-gen}$ is given by
\begin{align}
    &S_{\rm JT-gen} =
    \frac{1}{16\pi G_N} \int_{\mathcal{M}} {\rm d}z{\rm d}u\, \sqrt{-g}\, \phi\left(R+ \frac{2}{L^2}\right)+\frac{1}{8\pi G_N}\int_{\partial {\mathcal{M}}} {\rm d}u\,  \sqrt{-\gamma}\phi_b K\nonumber\\& \qquad\quad -\frac{1}{16\pi G_N}\int_{\mathcal{M}} {\rm d}z{\rm d}u\,  \sqrt{-g} \mathcal{L}_{\rm matter}[g,\chi] + S_{\rm c.t.}.
\end{align}
Above, $\phi_b$ is the boundary value of the dilaton, and $\mathcal{L}_{\rm matter}$ stands for the Lagrangian density of generic bulk matter $\chi$ (including the bulk gauge field) which couples only to the background metric but not to the dilaton.

Furthermore, the on-shell gravitational action is computed by imposing the following boundary conditions
\begin{equation}\label{Eq:phi-bc}
    \lim_{z\rightarrow 0}z \,\phi(z,u) = \phi_r(u), \quad \lim_{z\rightarrow 0}z^2 g_{uu}(z,u) = -1
\end{equation}
at the asymptotic boundary $z=0$. We note that we should further set $\phi_r(u)$, the non-normalizable mode of the dilaton, to the value $\phi_0$ as indicated in \eqref{Eq:ST-hol}. Here, it is understood that we choose the radial coordinate $z$ in which the metric has a double pole at the boundary. (We note that $z$ and the $\rho$ coordinate in Sec. \ref{Sec:GCM-deriv} are related via $\rho = 2 \sqrt{L z}$.) Furthermore, $S_{\rm JT}^{\rm on-shell}$ is obtained after removing the infrared divergences (mimicking the ultraviolet divergences of the dual theory) with necessary local counterterms in $S_{\rm c.t.}$. We will explicitly compute $S_{\rm JT}^{\rm on-shell}$ below. 

The general solution for the bulk metric which should satisfy $$R = - 2/L^2$$ (i.e. must be locally AdS$_2$) and satisfy the boundary condition \eqref{Eq:phi-bc} is:
\begin{align}\label{Eq:metricFG}
{\rm d}s^2 = \frac{L^2}{z^2}{\rm d}z^2 -L^2\left(\frac{1}{z}+ \frac{1}{2}z \, {\rm Sch}(f(u),u)\right)^2{\rm d}u^2.
\end{align}
in the Fefferman-Graham gauge where $g_{zz} = L^2/z^2$ and $g_{zu}=0$. (Clearly, the above is related to the metric \eqref{Eq:EMSol21} by $\rho = 2\sqrt{L z}$.) Let us first consider $\mathcal{L}_{\rm matter} =0$ for simplicity. The dilaton equations of motion $$\nabla_\mu\nabla_\nu\phi - g_{\mu\nu}\nabla^2\phi +\frac{1}{L^2}\phi =0$$ with the boundary condition \eqref{Eq:phi-bc} are satisfied if:
\begin{equation}\label{Eq:dilatonFG}
    \phi(z,u) = \frac{\phi_r(u)}{z} +z \zeta(u)
\end{equation}
and
\begin{align}
&\zeta(u) = -\frac{1}{2}\left(\phi_r''(u)+ \phi_r(u) {\rm Sch}(f(u),u)\right),\label{Eq:cons-phi1}\\
&\phi_r'''(u) + 2 \phi_r'(u) {\rm Sch}(f(u),u) + \phi_r(u) \frac{{\rm d}}{{\rm d}u}{\rm Sch}(f(u),u)= 0.\label{Eq:cons-phi2}
\end{align}
It turns out that it is sufficient to compute the renormalized on-shell action imposing \eqref{Eq:metricFG}, \eqref{Eq:dilatonFG} and \eqref{Eq:cons-phi1} only. As discussed below, the final constraint \eqref{Eq:cons-phi2} can be obtained directly from extremizing the renormalized on-shell action itself with respect to the time reparametrization mode $f(u)$ which is the only degree of freedom in the gravitational theory. 

In order to regularize the infrared divergence of the on-shell gravitational action, we need to choose the hypersurface $z = \epsilon$ where the induced metric is $\gamma_{uu} \sim - \epsilon^{-2}$ (note that $\epsilon$ is infinitesimal). Explicitly, the extrinsic curvature of the hypersurface $z = \epsilon$ in the ambient metric \eqref{Eq:metricFG} is given by
\begin{equation}
    K = \frac{1}{L}\left(1 - \frac{4}{2+ \epsilon^2 {\rm Sch}(f(u),u)}\right) = -\frac{1}{L} + {\rm Sch}(f(u),u)\epsilon^2 + \mathcal{O}(\epsilon^4).
\end{equation}
After choosing
\begin{equation}
    S_{\rm c.t.} =\frac{1}{8\pi G} \int {\rm d}u\, \sqrt{-\gamma}\,\phi_b(u) \times \frac{1}{L},
\end{equation}
imposing \eqref{Eq:metricFG}, \eqref{Eq:dilatonFG} and \eqref{Eq:cons-phi1}, and taking the limit $\epsilon\rightarrow 0$, we find that
\begin{align}\label{Eq:hol-ren}
    W_{\rm QT}[\phi_r(u)] = S_{\rm JT}^{\rm on-shell} = \frac{1}{8\pi G_N}\int {\rm d}u \,\phi_r(u) {\rm Sch}(f(u),u)
\end{align}
in agreement with \cite{Almheiri:2014cka,Maldacena:2016hyu,Engelsoy:2016xyb,Maldacena:2016upp,Joshi:2019wgi}. We can readily check that extremizing the above with respect to $f(u)$ yields the constraint \eqref{Eq:cons-phi2} as claimed above.\footnote{In the literature, the holographic renormalization has been discussed from a different point of view \cite{Almheiri:2014cka,Maldacena:2016hyu,Engelsoy:2016xyb,Maldacena:2016upp,Joshi:2019wgi} where the usual metric ${\rm d}s^2 = (L^2/z^2) ({\rm d}z^2 - {\rm d}t^2)$ is chosen along with a general cut-off hypersurface $z(u) = \epsilon t'(u)$ where the induced metric is $\gamma_{tt} = - 1/\epsilon^2$. The $\epsilon\rightarrow 0$ limit reproduces \eqref{Eq:hol-ren}. It has been more convenient for us to take the \textit{passive} view where the metric is written after time reparametrization and we do a more conventional holographic renormalization by choosing $z = \epsilon$ cutoff hypersurface. In either case, \eqref{Eq:hol-ren} is obtained as the regularization breaks the time reparametrization symmetry to SL(2,R).}

Therefore, the full action $S_{\rm T}$ given by \eqref{Eq:ST-hol} in absence of bulk matter fields simply reduces to
\begin{eqnarray}\label{Eq:FullAction1}
    S_{\rm T} &=& S_{{\rm GCM}} + \frac{\phi_0}{8\pi G_N}\int {\rm d}u \, {\rm Sch}(f(u),u)\nonumber\\
    &=& \int {\rm d}u \left(\frac{1}{2}x'(u)^2 - \frac{q}{2 x(u)^2}-\left(\frac{1}{4}x(u)^2-\frac{ \phi_0}{8\pi G_N}\right)\,{\rm Sch }(f(u),u) \right).
\end{eqnarray}
The above action provides the equations of motion for the two degrees of freedom, namely $x(u)$ and $f(u)$. As shown in Sec. \ref{Sec:GCM-action}, the equation of motion for $f(u)$ coming from $S_{{\rm GCM}}$ remarkably vanishes when $x(u)$ is on-shell. Therefore, the equation of motion for $f(u)$ is simply
\begin{equation}\label{Eq:Schcons}
    \frac{{\rm d}}{{\rm d}u}{\rm Sch}(f(u),u) =0.
\end{equation}
The above is just a special case of \eqref{Eq:cons-phi2}. Of course, the equation of motion of $x(u)$ is \eqref{Eq:x-Eom1} as in generalized conformal mechanics. The upshot is that the particle does not backreact on the time-reparametrization mode (and thus the bulk metric) and also on the dilaton.

These results can be readily generalized in presence of bulk matter fields. In the presence of bulk matter, the dilaton takes the form \cite{Joshi:2019wgi}
\begin{equation}\label{Eq:dilatonFGn}
    \phi(z,u) = \frac{\phi_r(u)}{z} +z \zeta(u) + S(z,u)
\end{equation}
where $S(z,u)$ is determined by bulk matter, and the constraints \eqref{Eq:cons-phi1} and \eqref{Eq:cons-phi2} are modified to 
\begin{align}
&\zeta(u) = -\frac{1}{2}\left(\phi_r''(u)+ \phi_r(u) {\rm Sch}(f(u),u) + \lim_{z\rightarrow 0}z T_{zz}(z,u)\right),\label{Eq:cons-phi1n}\\
&\phi_r'''(u) + 2 \phi_r'(u) {\rm Sch}(f(u),u) + \phi_r(u) \frac{{\rm d}}{{\rm d}u}{\rm Sch}(f(u),u)\nonumber\\&=   \lim_{z \rightarrow0}\left( T_{zu}(z, u) - z\partial_u T_{zz}(z, u) \right),\label{Eq:cons-phi2n}
\end{align}
where $T_{\mu\nu}$ is the energy-momentum tensor of the bulk matter. The final constraint \eqref{Eq:cons-phi2n} can be obtained from the extremization of the renormalized on-shell gravitational action (obtained by imposing \eqref{Eq:metricFG}, \eqref{Eq:dilatonFGn}, \eqref{Eq:cons-phi1n} and the matter equations of motion) with respect to $f(u)$. Explicitly, the full action is
\begin{equation}
    S_{\rm T} = S_{{\rm GCM}} + \frac{\phi_0}{8\pi G_N}\int {\rm d}u \, {\rm Sch}(f(u),u) + \,\, {\rm bulk\,\, matter\,\, contributions}.
\end{equation}
The equation of motion of $f(u)$ in this case is
\begin{align}\label{Eq:Eom-t-1}
      \phi_0 \frac{{\rm d} }{{\rm d}u}{\rm Sch} (f(u),u)
     =  \lim_{z \rightarrow0}\left( T_{zu}(z, u) - z\partial_u T_{zz}(z, u) \right).
\end{align}
Therefore, even in the presence of bulk matter fields, the backreaction of $x(u)$ on the time-reparametrization mode $f(u)$ (and thus the bulk metric), the dilaton and also bulk matter fields vanishes. The time reparametrization mode $f(u)$ is determined only by the bulk matter fields. 

We also note that the Hamiltonian of the full theory is simply
\begin{equation}
    H_{\rm T} = \mathbb{H} + H_{\rm grav},
\end{equation}
where $\mathbb{H}$ is the Hamiltonian of the generalized conformal mechanics which takes the form \eqref{Eq:Hniceform} and $H_{\rm grav}$ is the Hamiltonian of the gravitational theory. This split property of the Hamiltonian leads to the interpretation that the generalized conformal mechanics is a hair like degree of freedom \cite{Kibe:2023ixa}.

It is not hard to generalize to the case where we couple the quantized generalized quantum conformal mechanics to classical JT gravity. Following \cite{Kibe:2023ixa}, we should replace the $S_{\rm GCM}$ with $S_{\rm GCQM}$ given by 
\begin{equation}\label{Eq:SCQMQ}
    S_{\rm GCQM} = \int {\rm d}u\, \bra{\psi}\left(i\partial_u - \hat{\mathbb{H}}\right)\ket{\psi}, \quad \hat{\mathbb{H}}= \frac{1}{2}\hat{p}^2 + \frac{q}{2\hat{x}^2} +\frac{1}{4} {\rm Sch}(f(u),u) \hat{x}^2.
\end{equation}
$S_{\rm GCQM}$ should be extremized with respect to the state $\ket{\psi}$ simply yielding the Schr\"{o}dinger equation which explicitly is
\begin{equation}\label{Eq:Sch}
    i\partial_u \psi(x,u) = \left(-\frac{1}{2} \frac{\partial^2}{\partial x^2} + \frac{q}{2 x^2} + \frac{1}{4} {\rm Sch}(f(u),u)x^2 \right)\psi(x,u)
\end{equation}
in the position basis. The full action is
\begin{equation}
    S_{\rm T} = S_{{\rm GCQM}} + \frac{\phi_0}{8\pi G_N}\int {\rm d}u \, {\rm Sch}(f(u),u) + \,\, {\rm bulk\,\, matter\,\, contributions}.
\end{equation}
The equation of motion for $f(u)$ is 
\begin{align}\label{Eq:cons-phi2n2}
  &2\pi G_N \frac{{\rm d}^3}{{\rm d}u^3} \langle \hat{x}^2(u)\rangle+  4\pi G_N \frac{{\rm d}}{{\rm d}u} \langle \hat{x}^2(u)\rangle {\rm Sch}(f(u),u) + \left(2\pi G_N\langle \hat{x}^2(u) \rangle+ \phi_0\right)\frac{{\rm d}}{{\rm d}u}{\rm Sch}(f(u),u)\nonumber\\&=   \lim_{z \rightarrow0}\left( T_{zu}(z, u) - z\partial_u T_{zz}(z, u) \right).
\end{align}

In the following section, we will obtain the general normalizable solutions of the Schr\"{o}dinger equation satisfying appropriate boundary conditions. We will show that, although $\langle \hat{x}(u) \rangle$ is generically different from the solution of the classical equation of motion, generally 
\begin{equation}\label{Eq:Qx2}
  \langle \hat{x}^2(u)\rangle =  \frac{a_1 + a_2 f(u)+ a_3 f(u)^2}{f'(u)},  
\end{equation}
in an arbitrary state satisfying the Schr\"{o}dinger equation with $a_1$, $a_2$ and $a_3$ being (state dependent) constants. Recall that for a solution of the classical equation of motion we require that $a_1 a_3 - (a_2^2/4) = q$ as should be evident from \eqref{Eq:xSol1}. However, in the quantum theory, $a_1$, $a_2$ and $a_3$ are independent. 

As should be clear from the results of Sec. \ref{Sec:GCM-deriv} (recall \eqref{phi-sol}), the constraints for the dilaton (and thus the equation of motion for $f(u)$) are automatically satisfied in pure JT gravity whenever $\phi_r(u)$ assumes the \textit{general} form given by \eqref{Eq:Qx2} for arbitrary values of $a_1$, $a_2$ and $a_3$. It can also be checked by substituting \eqref{Eq:Qx2} in \eqref{Eq:cons-phi2n2} that the latter simply reduces to \eqref{Eq:Eom-t-1} in the presence of bulk matter fields implying that the backreaction of the quantum particle on the time-reparametrization mode (and thus the bulk metric), the dilaton and bulk matter fields vanish identically as in the case of the classical particle.

In the following section, we will explicitly study the normalizable solutions of the Schr\"{o}dinger equation and discuss how the time-reparametrization mode of the gravitational theory can be detected via suitable measurements.

We conclude this section with two remarks. Firstly, the backreaction of the particle on the gravitational sector does not vanish away from the large $N$ limit. Secondly, as discussed the alternative reformulation of the action presented in Appendix \ref{Sec:Full2} not only identifies the $f(u)$ in the action of the particle with the time reparametrization mode of JT gravity automatically, but it is also more suitable for quantization.

\section{Quantization of generalized conformal mechanics and the full semiclassical theory}\label{Sec:QGCM}

\subsection{The normalizable solutions of the Schr\"{o}dinger equation}

Our first aim is to obtain the normalizable solutions of the Sch\"{o}dinger equation \eqref{Eq:Sch} for generalized conformal mechanics. This cannot be approached by solving the spectrum of the Hamiltonian as the Hamiltonian does not have normalizable eigenstates (including of course ground state) when ${\rm Sch}(f(u),u)\leq 0$. The case  ${\rm Sch}(f(u),u)= 0$ is identical to usual quantum conformal mechanics analyzed in de Alfaro, Fubini and Furlan \cite{deAlfaro:1976vlx}. Since, the mass of the black hole in the JT gravity is $M_{BH}\sim -2\,{\rm Sch}(f(u),u) >0$, the Hamiltonian of the quantized generalized conformal mechanics has no normalizable eigenstate whenever the time reparametrization mode corresponds to a dynamical black hole solution of JT gravity. In what follows, we will assume that ${\rm Sch}(f(u),u) <0$ and $f'(u) >0$ for $u \in (-\infty, \infty)$.

It is easy to see this in the case of the usual conformal mechanics (with $f(u) =u$), for which the Hamiltonian is $H_0$ (given by \eqref{Eq:h0d0k0}).
The commutation relation \eqref{Eq:Alg1} implies that if $\ket{E}$ is any finite energy eigenstate with energy $E$, then $e^{i\alpha D_0}$ is also an energy eigenstate with energy $e^{2\alpha} E$ \cite{Britto-Pacumio:1999dnb}. Thus the spectrum is continuous and the eigenstates are non-normalizable in this case. The case $q=0$ is, in fact, just the free particle. 

For a more general argument, consider 
\begin{equation}
    O = \mu_O H+ \nu_O D + \omega_O K = \tilde{\mu}_O \tilde{H} + \tilde{\nu}_O \tilde{D} + \tilde{\omega}_O\tilde{K},
\end{equation}
with both $\{H, D, K\}$, and $\{ \tilde{H},\tilde{D},\tilde{K} \}$  satisfying the sl(2,R) Lie algebra. Then $$\Delta_O=\nu_O^2 - \mu_O\omega_O = \tilde{\nu}_O^2 - \tilde{\mu}_O\tilde{\omega}_O$$and $O$ has a discrete spectrum with normalizable eigenstates iff \cite{deAlfaro:1976vlx} $$\Delta < 0.$$ The Hamiltonian \eqref{Eq:HamilN} of generalized conformal mechanics can be written as
\begin{equation}
    \mathbb{H}  = \mathcal{Q}\bullet \mathcal{G} = H_0+\frac{1}{2}{\rm Sch }(f(u),u)\, K_0.
\end{equation}
Above we have used \eqref{Eq:Hniceform} (with $\mathcal{Q} =\{Q_0(u), Q_1(u), Q_2(u)\}$, the conserved charges of the Schwarzian theory and $\mathcal{G} =\{H_f, D_f, K_f\}$, the conserved charges of generalized conformal mechanics) and \eqref{Eq:h0d0k0} (we have ignored the term proportional to identity in \eqref{Eq:HamilN}). Recalling (from \eqref{Eq:Alg1}) that $H_0$, $D_0$ and $K_0$ furnish a representation of the sl(2,R) Lie algebra and \eqref{Eq:Q2}, we note that $\Delta_{\mathbb{H}}$ corresponding to the Hamiltonian is given by $$\Delta_{\mathbb{H}}=\frac{1}{4}\mathcal{Q}\bullet\mathcal{Q} =- \frac{1}{2}{\rm Sch }(f(u),u).$$ Therefore, $\Delta < 0$ is not satisfied if ${\rm Sch }(f(u),u)\leq 0$, and then, as claimed above, $\mathbb{H}$ does not have normalizable eigenstates in this case.

It is therefore not appropriate to describe the Hilbert space using the energy eigenbasis. Following \cite{deAlfaro:1976vlx}, we will describe the Hilbert space using the eigenbasis of an operator that is a linear combination of the conserved charges for the action \eqref{Eq:SNCQM}, namely $H_f$, $D_f$ and $K_f$ defined in \eqref{Eq:Alg2}. A generic operator of this form is 
\begin{equation}\label{Eq:geninv}
    G = \mu_G H_f + \nu_G D_f + \omega_G K_f,
\end{equation}
where $\mu_G$, $\nu_G$ and $\omega_G$ are constants. $G$ has a discrete spectrum with normalizable ground states iff $$\Delta_G = \nu_G^2 - \mu_G \omega_G <0.$$It was suggested in \cite{deAlfaro:1976vlx} that such a $G$ should be thought of as the true Hamiltonian of the system (which generates flow along a reparametrized time). The definition of $G$ automatically introduces a length scale which leads to an energy gap and regulates the infrared problem of absence of normalizable eigenstates of the Hamiltonian (more below). This is similar to the case of confinement in gauge theories \cite{Gross:1973id,Politzer:1973fx} where an energy scale emerges by spontaneously breaking scale invariance. 

Crucially, we want to point out that instead of following \cite{deAlfaro:1976vlx}, we can use a conserved charge $G$ with a discrete spectrum and normalizable eigenstates to find normalizable solutions of the Schr\"{o}dinger equation \eqref{Eq:Sch} with the \textit{original} Hamiltonian. This method of solving the time-dependent Schr\"{o}dinger equation is known as the method of invariants \cite{PhysRevLett.18.510,choi2004coherent} and has been used earlier in \cite{Kibe:2023ixa} in a similar context. To see how this works, note that since $G$ is a conserved charge, it should satisfy
\begin{equation}\label{Eq:Hei}
    \partial_u G + i [\mathbb{H}, G] = 0
\end{equation}
in the Schr\'{o}dinger picture also. Let us denote $\ket{G_n}$ to be an eigenstate of $G$ with eigenvalue $g_n$. Note that since $G$ has explicit time-dependence, any eigenstate $\ket{G_n}$ is also explicitly time-dependent. However, it can be easily shown that any eigenvalue $g_n$ of $G$ is constant (time independent) as follows. Firstly, plugging the state $e^{i \xi_n(u)} \ket{G_n}$ into the time dependent Schr\"{o}dinger equation \eqref{Eq:Sch} we obtain
\begin{equation}
  \mathbb{H} e^{i \xi_n(u)} \ket{G_n}= i\partial_u( e^{i \xi_n(u)} \ket{G_n})= i \left(i e^{i \xi_n(u)} \xi_n'(u)\ket{G_n} + e^{i \xi_n(u)} \partial_u \ket{G_n}\right).
\end{equation}
Taking an inner product of the above with $e^{i \xi_n(u)} \ket{G_n}$ we obtain
\begin{equation} \label{Eq:phase}
    \xi_n'(u) = \braket{G_n | i \partial_u - \mathbb{H} | G_n}.
\end{equation}
Therefore, $e^{i \xi_n(u)} \ket{G_n}$ solves the Schr\"{o}dinger equation if $\xi_n(u)$ satisfies the above equation. Secondly, we can differentiate the identity
\begin{equation*}
    G e^{i\xi_n(u)}\ket{G_n} = g_n(u)  e^{i\xi_n(u)} \ket{G_n}
\end{equation*}
with respect to time (assuming that $g_n$ are time-dependent) to obtain
\begin{equation*}
    -i(\mathbb{H} G - G \mathbb{H}) e^{i\xi_n(u)}\ket{G_n} + G \partial_u (e^{i\xi_n(u)}\ket{G_n}) = g_n'(u) e^{i\xi_n(u)}\ket{G_n}+ g_n(u) \partial_u (e^{i\xi_n(u)}\ket{G_n}).
\end{equation*}
after utilizing \eqref{Eq:Hei}. As $e^{i\xi_n(u)}\ket{G_n}$ is a solution of the Schr\"{o}dinger equation, using the eigenvalue equation itself we readily obtain that 
\begin{equation}
    g_n'(u)= 0.
\end{equation}
Therefore, finding the normalizable solutions of the Schr\"{o}dinger equation amounts to finding the time-independent spectrum of the invariant $G$, the corresponding time-dependent normalizable eigenstates and the corresponding time-dependent phases via \eqref{Eq:phase}.

Following \cite{deAlfaro:1976vlx}, we can consider the following dimensionless operators
\begin{align}\label{Eq:RST}
    R &= \frac{1}{2}\left(\frac{K_f}{\gamma}+ \gamma H_f \right)\nonumber\\
    S&= \frac{1}{2}\left(\frac{K_f}{\gamma}-\gamma H_f \right)\nonumber\\
    T&=-\frac{1}{2} D_f,
\end{align}
where $\gamma$ is a constant with mass dimension $-1$.\footnote{Note that $H_f$, $D_f$ and $K_f$ have mass dimensions 1, 0 and -1, respectively. So $R$, $S$ and $D$ are dimensionless.} The above operators satisfy
\begin{equation}\label{Eq:Alg5}
     [S,R] = -i T,\quad [R,T] = -i S,\quad [S,T]= -i R,
\end{equation}
which is the $o(2,1)$ Lie algebra of non-compact rotations. Clearly $R$ is a suitable invariant as it has normalizable eigenstates ($\Delta_R = - 1/4 <0$). Thus we can reduce the problem of finding solutions of the Sch\"{o}dinger equation to finding the spectrum of $R$, and the corresponding (time-dependent) eigenstates and phases. The energy scale $\gamma^{-1}$ spontaneously breaks the classical scale invariance of generalized conformal mechanics.

Recall that the Casimir  of the sL(2,R) Lie algebra is
\begin{equation}
    \mathcal{C}=\tilde{H}_f \tilde{K}_f - \frac{1}{4} \tilde{D}_f^2 =R^2 - S^2 - T^2= \frac{q}{4} -\frac{3}{16}.
\end{equation}
We can rewrite the Casimir in terms of a new variable $r_0$ as
\begin{equation}\label{Eq:r0Cas}
    \mathcal{C} = \frac{q}{4} -\frac{3}{16} = r_0(r_0 -1).
\end{equation}
Solving this quadratic equation we get
\begin{equation}\label{Eq:r0q}
    r_0 = \frac{1}{2} \left(1 \pm \sqrt{q+\frac{1}{4}}\right).
\end{equation}
For reasons described below, we choose $r_0$ as the root with the $+$ sign. We note that we also require that $q \geq -1/4$ for $r_0$ to be real. However, we will see that we would require $q > 0$ (we will comment more on this later).  We can define raising and lowering operators as
\begin{equation}
    L_{\pm}^{R} = S \pm i T,
\end{equation}
which obey
\begin{equation}
    [R, L_{\pm}^R] = \pm L_{\pm}^R, \quad [L_+^R , L_-^R] = -2 R.
\end{equation}
The normalized eigenstates $\ket{r_0, r}$ of $R$ with eigenvalues $r$ ($R \ket{r_0, r} =  r\ket{r_0, r}$)  furnish an irreducible representation of SL(2,R) (we will discuss below that $r_0$ should satisfy an inequality). It follows from \eqref{Eq:Alg5} that
\begin{equation}
    L_{\pm}^R\ket{r_0,r} = c_{\pm}(r_0,r) \ket{r_0,r\pm 1},
\end{equation}
and using \eqref{Eq:r0Cas} and $\mathcal{C} = R^2 - L_+^RL_-^R$, one can readily verify that
\begin{equation}\label{Eq:cpm}
    |c_{\pm}(r_0,r)|^2 = r(r\pm1) - r_0(r_0-1) \geq 0,
\end{equation}
which implies that 
\begin{equation}
    r\geq r_0, \quad {\text or} \quad r \leq -r_0.
\end{equation}
Note also that we can choose the phases appropriately so that $$c_\pm(r_0,r)= \sqrt{r(r\pm1) - r_0(r_0-1)}.$$We readily note that  
\begin{equation}
    c_-(r_0, r_0) = 0, \quad {\rm and} \quad c_+(r_0, -r_0) = 0
\end{equation}
implying that   
\begin{equation}
    L_-^R \ket{r_0, r =r_0} = 0, \quad L_+^R \ket{r_0, r= - r_0} = 0.
\end{equation}
We thus obtain the discrete series of eigenstates $\ket{r_0, r_n}$ with eigenvalues
\begin{equation}
    r_n = r_0 + n, \quad n=0,1,2,\hdots
\end{equation}
by repeated action of $L_+^R$ on $\ket{r_0, r= r_0}$, and similarly another discrete series of eigenstates $\ket{r_0, \tilde{r}_n}$ eigenvalues $\tilde{r}_n = - r_0 - n$ by repeated action of $L_-^R$ on $\ket{r_0, r=-r_0}$. As should be clear from the discussion below, we should choose the series of eigenvalues with $r \geq r_0 > 3/4$. 

The eigenstate $\ket{r_0, r = r_0}$ with $R$ eigenvalue $r_0$ can be found simply by solving
\begin{equation}
    L_-^R \psi_0(x,u) =0,
\end{equation}
where  $\psi_0(x,u) = \bra{x} r_0, r=r_0\rangle$.  The above can be written as
\begin{equation}
    \left(\frac{K_f}{\gamma}-R+\frac{i}{2} D_f\right) \psi_0(x) = \left(K_f-r_0+\frac{i}{2} D_f\right) \psi_0(x)=0,
\end{equation}
The normalizable solution to this differential equation is (see Appendix \ref{App:wavefunc} for details)
\begin{multline}\label{Eq:psi0}
\psi_0(x,u) = \sqrt{\frac{2}{\Gamma(2 r_0)}}\left(\frac{\gamma f'(u)}{\gamma^2+f(u)^2}\right)^{r_0} \\x^{2 r_0 - \frac{1}{2}}\exp \left( -\frac{x^2 \gamma f'(u)}{2(\gamma^2+f(u)^2)} +i \frac{x^2}{4 f'(u)} \left(\frac{2 f'(u)^2 f(u)}{\gamma^2+f(u)^2} -f''(u)\right)\right),
\end{multline}
Note that this wavefunction is normalizable since $f'(u) > 0$.  

We recall that $x=0$ is the boundary of the AdS$_2$ space (as $x$ is essentially the radial coordinate). The wavefunction should vanish smoothly outside the throat, so we require that 
\begin{equation}
    \psi_0(0,u) = \partial_x \psi_0(0,u) = 0.
\end{equation}
As observed in \cite{deAlfaro:1976vlx} and can be readily seen from \eqref{Eq:psi0}, this requires that
\begin{equation}
    r_0 > \frac{3}{4},
\end{equation}
which corresponds to picking the positive root in \eqref{Eq:r0q} with
\begin{equation}
    q >0.
\end{equation}
As a result, $\mathcal{C}> - 3/16$. Furthermore, we are restricted to using only the positive eigenvalue series of $R$.

We recall from Sec. \ref{Sec:GCM-deriv} that when $q>0$, the particle is classically is not repelled away from the throat, and therefore it makes sense that the particle remains confined to the throat quantum mechanically also in this case.\footnote{It is to be noted that the quantization can make sense with other boundary conditions for arbitrary values of $q$. As discussed in \cite{Andrzejewski:2015jya}, quantization of conformal mechanics via geometric quantization of the Hamiltonian dynamics on the coadjoint orbit is possible for any value of $q$. Interestingly, there is a one parameter family of inequivalent quantizations where the Hamiltonian is self-adjoint when $-1/4 < q < 3/4$. It would be interesting to study these general possibilities for the generalized quantum mechanics for its intrinsic worth.}

The other eigenstates of $R$ can be generated by the repeated action of the raising operator $L_+^R$, as for instance the normalized first excited eigenstate of $R$ is 
\begin{equation}
    \psi_1(x,u) = \bra{x} r_0, r_0+1\rangle = \frac{1}{c_+(r_0,r_0)} L_+^R \psi_0(x,u)=\left(\frac{K_f}{\gamma}-r_0-\frac{i}{2} D_f\right) \psi_0(x,u).
\end{equation}
Explicitly,
\begin{multline}\label{Eq:psi1}
    \psi_1(x,u) = \frac{1}{\sqrt{r_0 \Gamma(2 r_0)}} \left(\frac{2\gamma f'(u)}{\gamma^2+f(u)^2}\right)^{r_0} \frac{2 r_0(\gamma^2+f(u)^2) - \gamma x^2 f'(u)}{(-i \gamma+f(u))^2}\\x^{2 r_0 - \frac{1}{2}}\exp \left( -\frac{x^2 \gamma f'(u)}{2(\gamma^2+f(u)^2)} +i \frac{x^2}{4 f'(u)} \left(\frac{2 f'(u)^2 f(u)}{\gamma^2+f(u)^2} -f''(u)\right)\right).
\end{multline}
Generally we can label the (time-dependent) normalized eigenstates of $R$ as $\ket{r_0, r_n,u} \equiv \ket{\psi_n(u)}$ with eigenvalues $r_n = r_0 + n$, respectively ($n= 0,1,\cdots$), and $\psi_n(x,u) = \bra{x}\psi_n(u)\rangle$ as the wavefunction in the position basis. Here and for rest of this section, we would keep the time dependence of the $R$-eigenstates $\ket{r_0, r_n,u} \equiv \ket{\psi_n(u)}$ explicit.

Finally, to find the solutions of the Schr\"{o}dinger equation \eqref{Eq:Sch}, we need to find the time-dependent phases $\xi_n(u)$ for each $\psi_n(x,u)$ via \eqref{Eq:phase}.  Remarkably, as shown in Appendix \ref{App:phase}, the phases are independent of $n$, i.e. the eigenstate, and are explicitly 
\begin{equation}\label{Eq:xi}
    \xi_n(u) = -2 r_0 \arctan\left(\frac{f(u)}{\gamma}\right) := \xi(u)
\end{equation}
Therefore, the most general normalized solution of the Schr\"{o}dinger equation \eqref{Eq:Sch} is
\begin{equation}
    \psi(x,u) = e^{i\xi(u)} \sum_{n=0}^\infty a_n \psi_n(x,u), \quad \sum_{n=0}^\infty \vert a_n\vert^2 =1
\end{equation}
where $\psi_n(x,u)$ are the normalized eigenstates of $R$. We can also write the above in the form
\begin{equation}\label{Eq:Sch-sol}
    \ket{\psi(u)} = e^{i\xi(u)} \sum_{n=0}^\infty a_n \ket{r_0, r_n,u} = e^{i\xi(u)} \sum_{n=0}^\infty a_n \ket{\psi_n(u)}.
\end{equation}
A consequence of \eqref{Eq:xi} is that we can ignore the phase $\xi(u)$ for the computation of the expectation values of operators for an arbitrary solution of the Schr\"{o}dinger equation. 

To compute the expectation value of the Hamiltonian in an arbitrary normalizable solution of the Schr\"{o}dinger equation, the following identity 
\begin{align}\label{Eq:Hniceform-2}
    \mathbb{H} &= \mathcal{Q}\bullet\mathcal{G} = \frac{1}{2}\left(Q_0(u)\gamma + \frac{Q_2(u)}{\gamma}\right) R +\frac{1}{2}\left(Q_0(u)\gamma - \frac{Q_2(u)}{\gamma}\right) S + \frac{1}{2}Q_1(u) T\nonumber\\
    &=\frac{1}{2}\left(Q_0(u)\gamma + \frac{Q_2(u)}{\gamma}\right) R +\frac{1}{4}\left(Q_0(u)\gamma - \frac{Q_2(u)}{\gamma} - i Q_1(u)\right) L_+^R \nonumber\\& \quad + \frac{1}{4}\left(Q_0(u)\gamma - \frac{Q_2(u)}{\gamma} + i Q_1(u)\right) L_-^R
\end{align}
is useful (and it can be readily verified using \eqref{Eq:RST}). For a general solution $\ket{\psi(u)}$ of the Schr\"{o}dinger equation \eqref{Eq:Sch-sol}, we obtain (with $\langle O(u)\rangle \equiv \braket{\psi(u)| x | \psi(u)}$)
\begin{align}\label{Eq:ABC}
    &\langle R\rangle = r_0 + \sum_{n=0}^\infty \vert a_n\vert^2 n : = A, \quad  \langle L_+^R\rangle= \sum_{n=0}^\infty a_n \bar{a}_{n+1} \sqrt{(1+n)(n+ 2r_0)}:= B + i C,\nonumber\\
    &\langle L_-^R\rangle = \sum_{n=1}^\infty a_n \bar{a}_{n-1} \sqrt{n(n-1+ 2r_0)} =  B - i C,
\end{align}
where $A$, $B$ and $C$ are real state-dependent constants. Of course, $\langle R\rangle$ and $\langle L_\pm^R\rangle$ are time independent since $R$ and $L_\pm^R$ are conserved charges. It follows from \eqref{Eq:Hniceform-2} that (with $\langle \mathbb{H}(u)\rangle \equiv \braket{\psi(u)| \mathcal{Q}\bullet\mathcal{G} | \psi(u)}$)
\begin{equation}\label{Eq:Hexp}
    \langle \mathbb{H}(u)\rangle =\frac{1}{2} A \left(Q_0(u)\gamma + \frac{Q_2(u)}{\gamma}\right)  +\frac{1}{2}B\left(Q_0(u)\gamma - \frac{Q_2(u)}{\gamma} \right)+ \frac{1}{2}C Q_1(u).
\end{equation}
Note that when $f(u)$ corresponds to a static black hole solution in JT gravity, ${\rm Sch}(f(u),u)$ is a constant, and therefore $Q_i(u)$ are also constants by virtue of the identities \eqref{Eq:QWI}. This implies that $\langle \mathbb{H}\rangle$ is constant whenever $f(u)$ corresponds to a static black hole solution in JT gravity. This is also what is implied by the Ward identity \eqref{Eq:HfWI}.

The expectation value of the position and momentum operators in the $\ket{\psi_0(u)}$ state for arbitrary $f(u)$ are given by
\begin{align}\label{Eq:xp1quant}
    \braket{\psi_0(u)| x | \psi_0(u)} &= \frac{\Gamma(2 r_0+\frac{1}{2})}{\Gamma(2 r_0)} \sqrt{\frac{\gamma^2+f(u)^2}{\gamma f'(u)}},\\
    \braket{\psi_0(u)| p | \psi_0(u)} &= \partial_u \braket{\psi_0(u) | x | \psi_0(u)}
\end{align}
Note that we cannot choose $c_1$ and $c_2$ in the general classical solution \eqref{Eq:xSol1} in any way to match with $\braket{\psi_0(u)| x | \psi_0(u)}$.

For a general solution of the Schr\"{o}dinger equation \eqref{Eq:Sch-sol}, as for instance for the state $\ket{\psi(u)} = e^{i\xi(u)}(a_0 \ket{\psi_0(u)}+a_1 \ket{\psi_1(u)})$, $\braket{\psi(u)| x | \psi(u)}$ is also not of the square-root form, i.e.
\begin{equation}
 \langle x(u)\rangle =   \braket{\psi(u)| x | \psi(u)} \neq \sqrt{\frac{a_1 + a_2 f(u) + a_3 f(u)^2}{f'(u)}}
\end{equation}
for any constants $a_1$, $a_2$ and $a_3$ as in the general solution \eqref{Eq:xSol1} of the classical equations of motion.

\subsection{Solutions of the full semi-classical theory}

We have discussed in Sec. \ref{Sec:FT} that in order to find the semi-classical solutions of the full theory, we need to study $\langle x^2(u)\rangle \equiv \braket{\psi(u)| x^2 | \psi(u)}$ for an arbitrary solution $\ket{\psi(u)}$ of the Schr\"{o}dinger equation given by \eqref{Eq:Sch-sol}.

From the identity \eqref{Eq:x2nice}, we can readily obtain using \eqref{Eq:RST} that\footnote{Note that this identity for $x^2$ is valid both in the Schr\"{o}dinger picture, (where it is a time-independent operator) and also in Heisenberg picture. The same is true for \eqref{Eq:Hniceform-2}. Note also that $R$ and $L_\pm^R$ have explicit time dependence and are conserved charges. These identities are computationally useful.}
\begin{equation}\label{Eq:xniceform-2}
    x^2 =2 \frac{\gamma^2+f(u)^2}{\gamma f'(u)} R + \frac{(\gamma+ i f(u))^2}{\gamma f'(u)} L_+^R + \frac{(\gamma-i f(u))^2}{\gamma f'(u)}L_-^R.
\end{equation}
Using the definitions of $A$, $B$ and $C$ as given in \eqref{Eq:ABC}, we then obtain that
\begin{align}\label{Eq:x2exp}
    \langle x^2(u) \rangle  &= 2 A \frac{\gamma^2+f(u)^2}{\gamma f'(u)} + 2 B \frac{\gamma^2- f(u)^2}{\gamma f'(u)} - 4 C \frac{f(u)}{f'(u)} \nonumber\\
    &= \frac{2(A+B)\gamma^2 - 4 C \gamma f(u) +2 (A-B)f(u)^2}{\gamma f'(u)}
\end{align}
Taking the total time-derivative of both sides of \eqref{Eq:Hexp} and using the above, \eqref{Eq:QWI}, and that $R$ and $L_\pm^R$ are conserved charges, we can readily verify that the Ward identity \eqref{Eq:HfWI} is satisfied.\footnote{Since \eqref{Eq:Hniceform-2} and \eqref{Eq:xniceform-2} are valid in Heisenberg picture also, we can check the Ward identity \eqref{Eq:HfWI} directly as an operator identity as well using \eqref{Eq:QWI}, and that $R$ and $L_\pm^R$ are conserved charges.}

We also readily see from \eqref{Eq:x2exp} that $\langle x^2(u) \rangle$ is of the form \eqref{Eq:Qx2} with
\begin{equation}
    a_1 = 2(A+B)\gamma, \quad a_2 = -4 C, \quad a_3 = \frac{2(A-B)}{\gamma}.
\end{equation}
Therefore, the backreaction of the particle on the time reparametrization mode, the bulk metric, the dilaton and the bulk matter fields should vanish as discussed in Sec. \ref{Sec:FT} for an arbitrary normalizable solution of the Schr\"{o}dinger equation. The bulk metric, the dilaton and the bulk matter fields are determined just by the gravitational equations with the time reparametrization mode determined by \eqref{Eq:Eom-t-1} in the full theory.

There is one important thing to note. In order that, $A$, $B$ and $C$ are well defined, we require that the infinite sums in \eqref{Eq:ABC} are convergent. This is possible if $a_n$ decays faster than $1/n$ as $n\rightarrow\infty$. Therefore, the semi-classical Hilbert space is actually more restrictive than just the requirement of normalizability of the state. 

It is worth mentioning that there is also a nice geometric interpretation for the general solution of the Schr\"{o}dinger equation \eqref{Eq:Sch-sol} which involves the choice of the conserved charge $R$ which has a discrete spectrum and normalizable eigenstates. For any conserved charge $G$ of the generalized conformal mechanics, one can associate a Killing vector in the background metric \eqref{Eq:metricFG} which generates the same $SL(2,R)$ transformation. We can check that if $\Delta_{G} \geq 0$, then the corresponding bulk Killing vector becomes null in the domain $0< z < \infty$, and if $\Delta_{G} < 0$, then the corresponding bulk Killing vector is always time-like in the domain $0< z < \infty$. The general solution of the Schr\"{o}dinger equation \eqref{Eq:Sch-sol} can be interpreted as a evolution in the bulk (since $z$ is essentially $x$) in a time set by the time-like Killing vector corresponding to the conserved charge $R$. A similar observation has been made also in the context of usual conformal mechanics in \cite{Claus:1998ts}.

\subsection{Detecting the time reparametrization mode}\label{Sec:Detect}
The observables of the generalized quantum conformal mechanics of the particle coupled to JT gravity theory, like the Hamiltonian which takes the form \eqref{Eq:Hniceform}, depends explicitly on the bulk SL(2,R) frame, i.e. the values of the charges $\mathcal{Q}$ given by \eqref{Eq:Q}. This implies that we can reconstruct the time reparametrization mode of the JT gravity theory explicitly by performing suitable measurements on the particle.\footnote{As noted in Sec. \ref{Sec:GCM-deriv}, any dilaton configuration breaks the SL(2,R) isometry of the AdS$_2$ background to U(1).}

Suitable observables for measurements are $\langle \mathbb{H}(u)\rangle$ and $\langle x^2(u) \rangle$ which takes the forms \eqref{Eq:Hexp} and \eqref{Eq:x2exp}, respectively, in an arbitrary solution $\ket{\psi(u)}$ of the Schr\"{o}dinger equation (see \eqref{Eq:Sch-sol}) with the real constants $A$, $B$ and $C$ as defined in \eqref{Eq:ABC}. For a general function $f(u)$, we will need continuous weak measurements of $x^2$ and the energy to determine the function $f(u)$. In practice we will be able to attain only an approximation of the function $f(u)$ with continuous weak measurements \cite{Jacobs2006} of $\langle x^2(u) \rangle$ and the energy (which are fraught with inherent quantum uncertainties as discussed below). Furthermore, as $x^2$ and $\mathbb{H}$ do not commute, we will need to prepare the state twice to measure $\langle \mathbb{H}(u)\rangle$ and $\langle x^2(u) \rangle$. It is to be noted that the choice of $R$ (and thus $\gamma$) is actually up to the experimentalist. The choice of $R$ gives the basis in which the general solution $\ket{\psi(u)}$ of the Schr\"{o}dinger equation has been expressed in \eqref{Eq:Sch-sol}, and also for the definitions of $A$, $B$ and $C$ in \eqref{Eq:ABC}.

The procedure to extract the time reparametrization mode is as follows. Firstly, as observed before, from $\langle \mathbb{H}(u)\rangle$ and $\langle x^2(u) \rangle$ given by \eqref{Eq:Hexp} and \eqref{Eq:x2exp}, respectively, we can extract the identity
\begin{equation}
    \frac{{\rm d}}{{\rm d}u}{\rm Sch}(f(u),u) = \frac{4}{\langle x^2(u) \rangle}\frac{{\rm d}\langle \mathbb{H}(u)\rangle}{{\rm d}u}
\end{equation}
which is just the Ward identity \eqref{Eq:HfWI}. Regarding the above as a fourth order differential equation for $f(u)$, we can extract $f(u)$ up to four integration constants from $\langle x^2(u) \rangle$ and $\langle \mathbb{H}(u)\rangle$. As for instance, if the Schwarzian derivative of $f(u)$ is a constant, then
\begin{equation}
    f(u) =\beta \frac{a\tanh\left(\frac{\pi u}{\beta}\right) +b}{c \tanh\left(\frac{\pi u}{\beta}\right)+d}, \quad a d- b c= 1,
\end{equation}
where $a$, $b$, $c$ and $\beta$ are the four integration constants (note $d = (1 + bc)/a$). Note that $\beta$ is just the inverse temperature of a static black hole solution in JT gravity (with mass $4\pi^2/\beta^2$ since ${\rm Sch}(f(u),u) = - 2\pi^2/\beta^2$) in this case. In this case,
\begin{align}
 \langle x^2(u)\rangle &=\frac{2 \gamma}{\pi} \left(c \sinh\left(\frac{\pi u}{\beta}\right)+d \cosh \left(\frac{\pi u}{\beta}\right)\right)^2\nonumber \\
  &\Bigg(A \left(1 +\frac{\beta^2}{\gamma^2} \frac{\left(a\tanh\left(\frac{\pi u}{\beta}\right) +b\right)^2}{\left(c \tanh\left(\frac{\pi u}{\beta}\right)+d\right)^2}\right)+ B\left(1 -\frac{\beta^2}{\gamma^2} \frac{\left(a\tanh\left(\frac{\pi u}{\beta}\right) +b\right)^2}{\left(c \tanh\left(\frac{\pi u}{\beta}\right)+d\right)^2}\right)\\
  &\qquad- 2 C\frac{\beta}{\gamma} \frac{a\tanh\left(\frac{\pi u}{\beta}\right) +b}{c \tanh\left(\frac{\pi u}{\beta}\right)+d} \Bigg) \nonumber
\end{align}
as a special case of \eqref{Eq:x2exp}. In this case, we can extract the seven unknown real constants, namely $a$, $b$, $c$, $\beta/\gamma$, $A$, $B$ and $C$ by fitting $\langle x^2(u) \rangle$ with the above form. More generally, substituting $f(u)$ known up to four integration constants in \eqref{Eq:x2exp}, we can obtain a functional form of $\langle x^2(u) \rangle$ where we we have seven unknown real constants including $A$, $B$ and $C$. These seven constants should be determined simply by fitting with measured $\langle x^2(u) \rangle$. Thus we can reconstruct the time reparametrization mode $f(u)$ explicitly from the continuous weak measurements of $\langle \mathbb{H}(u)\rangle$ and $\langle x^2(u) \rangle$.

It is to be noted that the continuous weak measurements \cite{Jacobs2006} are fraught with quantum uncertainties. As for instance, the Heisenberg picture operators $x^2(0)$ and $x^2(u)$ do not commute. In fact, since in the Heisenberg picture, $x(u) \sim x(0) + p(0)u$ for small $u$, the commutation $[x^2(0), x^2(u)] \propto u$. Therefore, we need to do continuous weak measurement of $\langle x^2(u) \rangle$ for a sufficiently small time interval. Similarly, we should perform continuous weak measurement of $\langle \mathbb{H}(u) \rangle$ also for a sufficiently short time interval. These issues are dealt with in the standard theory \cite{Jacobs2006} of continuous weak measurement. 

In the large $N$ limit, the backreaction of the particle on the JT gravitational sector vanishes. Therefore, we do not need to take into account the gravitational degrees of freedom (including the time reparametrization mode) in the standard theory of continuous weak measurement as it is sufficient to just deal with the particle. However, away from the large $N$ limit, the backreaction of the particle on the gravitational sector and the time reparametrization mode does not vanish. Therefore, we need to develop a more sophisticated theory of measurement away from the large $N$ limit. We leave this for the future.

\section{Discussion}\label{Sec:Disc}
We have shown that generalized conformal quantum mechanics coupled to JT gravity naturally arises from considering a probe particle in the near-horizon geometry of a nearly extremal black hole, and leads to a solvable model for quantum gravity coupled with a detector. The full model is amenable to quantization, and can lead to a detailed understanding of how semi-classical spacetimes including black holes with entropies emerge from the point of view of a measuring device. In the future, we would like to study the full quantum theory of generalized conformal mechanics coupled to JT gravity, and explicitly understand the emergence of semi-classical spacetimes applying the theory of measurements.

Here, we discuss that our model can be possibly extended to higher dimensions leading to tractable quantizable models for black hole microstates. We first note that $n$-particle conformal mechanics arises from solutions in five-dimensional $\mathcal{N}=1$ supergravity, as for instance, containing $n$ static extremal black holes \cite{Michelson:1999zf,Britto-Pacumio:1999dnb}. In such solutions, the centers of the throats of the extremal black holes are vectors $\vec{x}_A$ in R$^4$ (with $A = 1,2, \cdots, n$), and the full five-dimensional geometry reduces to AdS$_2$ $\times$ S$^3$ when zoomed near $\vec{x}_A$ with the latter denoting the boundaries of the AdS$_2$ factors. When $\vec{x}_A$ become slowly varying functions of the time coordinate only, the supergravity action reduces to a conformal mechanics action of $n$-particles on R$^4$ at the leading order. It would be interesting to study the case of $n$ nearly extremal black hole solutions. We can expect that the conformal mechanics of the $n$ particles is generalized via coupling to the time reparametrization modes of the $n$ throats. With a single throat, the setup should reduce to the case studied in this work.

Another way of viewing the system of $n$ nearly extremal throats whose centers slowly vary with time would be to consider $n$ quantum dots whose motion in R$^4$ is governed by conformal quantum mechanics. Each quantum dot has internal degrees of freedom whose dynamics can be captured holographically by an emergent AdS$_2$ throat. Furthermore, the conformal mechanics couples to the time reparametrization modes of each AdS$_2$ throat. This can be called a \textit{fragmented} holographic setup since the emergent geometries (AdS$_2$ throats) capture only parts of the degrees of freedom (the internal degrees of freedom of the quantum dots at $\vec{x}_A$). If the AdS$_2$ throats are described by JT gravity, then the full theory can be amenable to quantization.

Motivated by the fragmentation instability  \cite{Maldacena:1998uz} of near horizon geometries of near extremal black holes in $D+1$ dimensions, we can view such a setup of JT gravity theories of $n$ two-dimensional throats (dual to the quantum dots) coupled to generalized conformal quantum mechanics of the motion of the center of the throats in R$^D$ (positions of the dual quantum dots) as a microstate model for such black holes. If the black hole has a horizon with radius $r_h$, then the motion of the throats are restricted to $\vert\vec{x}_A \vert\leq r_h$ (see \cite{Britto-Pacumio:1999dnb} for similar discussion). Quantum-mechanically, the multi-particle wavefunction should vanish whenever $\vert\vec{x}_A \vert = r_h$. 

Such microstate models of black holes inspired by the fragmentation instability \cite{Maldacena:1998uz} of near horizon geometries of nearly extremal black holes has been studied in \cite{PhysRevD.102.086008,Kibe:2021gtw,Kibe:2023ixa}. In this model the AdS$_2$ throats were restricted to end on a lattice which discretizes the horizon, and the AdS$_2$ throats were coupled by gravitational hair. The microstates are identified with the stationary solutions. It was shown that the relaxation and absorption properties of the microstates mimic those of a classical black hole. When quantum oscillators are coupled to one (or more throats), then they lose most of the energy, decouple from the throats and follow a quantum trajectory which retains a non-isometric copy of their initial state while another copy of the information of their initial state is transferred to the non-linear ringdown of the microstate as it transits to another \cite{Kibe:2023ixa}. Thus the model demonstrates replication of quantum information without paradoxes in a local semi-classical approximation realizing the black hole complementarity principle \cite{PhysRevD.48.3743,PhysRevD.50.2700,Harlow:2014yka,Raju:2020smc,Kibe:2021gtw} microscopically. Furthermore, the model explicitly demonstrates quantum information mirroring \cite{Hayden_2007} in consistency with the complementarity principle in which the explicit Hayden-Preskill decoding protocol of the infalling qubit from decoupled hair radiation does not depend on the state of the interior and  only very limited information of the pre-existing hair radiation \cite{PhysRevD.102.086008,Kibe:2023ixa}. When the throats Hawking radiate after coupled to baths, the throats homogenize at very long time with the information of the interior transferred to the Hawking radiation and the decoupled hair radiation. 

We can therefore expect that setups with JT gravity theories of $n$ throats coupled to generalized conformal quantum mechanics of the motion of the center of the throats in R$^D$, which are amenable to full quantization, can give deeper insights into the microscopic realization of the black hole complementarity principle without paradoxes, quantum information processing in black holes and also the emergence of the semi-classical geometries of black holes. In such models, the generalized conformal mechanics should play the role of the hair degrees of freedom and act as models of detectors for the interior of the quantum black hole.

\acknowledgments{AM acknowledges support from Fondecyt Grant 1240955.}
\appendix
\section{An alternative form of the full action}\label{Sec:Full2}
We can construct the full theory also using \textit{democratic coupling via auxiliary fields} as employed in \cite{Banerjee:2017ozx,Kurkela:2018dku,Ecker:2018ucc,Mondkar:2021qsf,Kibe:2023ixa}.\footnote{The principle of this coupling is to promote the marginal and relevant couplings of the two theories (describing two sectors such as perturbative and non-perturbative sectors) to auxiliary fields. These auxiliary fields are coupled via an ultralocal term. It is important that the extremization of the full action with respect to the auxiliary fields yields algebraic equations only. Solving the latter, we obtain the couplings (which have been promoted to auxiliary fields) of each sector in terms of expectation values of operators in the other sector. The full dynamics should be solved self-consistently. This coupling scheme has been originally proposed to provide semi-holographic effective descriptions of systems such as the quark-gluon plasma in consistency with the Wilsonian renormalization approach \cite{Banerjee:2017ozx,Kurkela:2018dku,Ecker:2018ucc,Mondkar:2021qsf}. However, it has been utilized also in building more general effective theories coupling multiple holographic systems as in \cite{Kibe:2023ixa} for constructing a black hole microstate model.}  The appropriate full action with auxiliary variables $J_1(u)$ and $J_2(u)$ with mass dimensions $2$ and $-1$ respectively, is
\begin{eqnarray}\label{Eq:CoupledAction0}
    S_{\rm T} &=& S_{CM}[J_1(u)] + W_{\rm QT}\left[\phi_r(u) = \phi_0-J_2(u)\right] \nonumber\\&&+\frac{1}{4\pi G_N} \int {\rm d}u \, J_1(u) J_2(u)   
\end{eqnarray}
where
\begin{equation}\label{Eq:SCMc}
    S_{\rm CM}[J_1(u)] = \int {\rm d}u\, \left(\frac{1}{2}x'(u)^2  -\frac{q}{2 x(u)^2} -\frac{1}{2}\, x(u)^2 J_1(u)\right),
\end{equation}
$G_N$ is the two dimensional Newton's constant, and $\phi_0$ is an arbitrary dimensionless constant. Here, $W_{\rm QT}$ is the logarithm of the partition function (quantum effective action) of a putative large N quantum dot theory dual to a two dimensional gravity theory, where $N^2\sim G_N^{-1}$. 

In the absence of bulk matter fields \eqref{Eq:CoupledAction0} the full action reduces to
\begin{equation}\label{Eq:CoupledAction1}
    S_{\rm T} = S_{\rm CM}[J_1(u)] +\frac{1}{8\pi G_N}\int {\rm d}u\, \left(\phi_0 - J_2(u)\right) {\rm Sch}(t(u),u)+ \frac{1}{4\pi G_N}\int {\rm d}u \, J_1(u) J_2(u)
\end{equation}
following the discussion in  Sec. \ref{Sec:FT}. Extremizing the above action w.r.t. $J_1$ and $J_2$ we obtain
\begin{equation}\label{Eq:J1J2}
    J_1(u) = \frac{1}{2} {\rm Sch}(t(u),u), \quad J_2(u) = 2 \pi G_N\, x^2(u).
\end{equation}
The equation of motion for $x(u)$ is
\begin{equation}\label{Eq:x-Eom0}
    x''(u)  = \frac{q}{x(u)^3} -   J_1(u)\, x(u),
\end{equation}
and the equation of motion for $f(u)$ is:
\begin{equation}\label{Eq:Eom-t-1-1}
     \left(J_2(u)+  \phi_0\right) \frac{{\rm d}}{{\rm d}u}{\rm Sch} (f(u),u) + 2 J_2'(u){\rm Sch}(f(u),u)+ J_2'''(u)  =0.
\end{equation}

Clearly when $J_1$ takes its on-shell value as given by \eqref{Eq:J1J2}, the equation of motion for $x(u)$ given by \eqref{Eq:x-Eom0} is identical to \eqref{Eq:x-Eom1} as obtained in generalized conformal mechanics. When $J_2(u)$ takes its the on-shell value of $2\pi G_N\, x^2(u)$ as given by \eqref{Eq:J1J2} and furthermore when $x(u)$ is on-shell, \eqref{Eq:Eom-t-1-1} reduces to simply \eqref{Eq:Schcons}. So we reproduce the equations of motion for $x(u)$ and $f(u)$ in the full theory in the earlier formulation.

Furthermore, when $J_1$ and $J_2$ are on-shell, clearly \eqref{Eq:CoupledAction1} reduces to \eqref{Eq:FullAction1} in absence of bulk matter fields. Therefore, we should indeed reproduce identical equations of motion for $x(u)$ and $f(u)$ in the full theory in the two formulations. This generalizes when we include bulk matter fields and quantize the generalized conformal mechanics.

There is one subtle distinction which can be observed by comparing \eqref{Eq:CoupledAction0} with \eqref{Eq:ST-hol}. The non-normalizable modes of the bulk dilaton are different in the two formulations. In this alternative formulation, although the equations of motion for $x(u)$ and $f(u)$ are identical to the older formulation in Sec. \ref{Sec:FT}, the bulk dilaton is determined here together by  $x(u)$ and $f(u)$ instead of just by $f(u)$. We can take the point of view that $f(u)$ and not the bulk dilaton is a real degree of freedom, and therefore at least in the large $N$ limit, the two formulations are equivalent. 

The alternative formulation is easier to quantize as the Schwarzian derivative of $f(u)$ does not directly enter in the action.

\section{Finding eigenstates of R}
\label{App:wavefunc}
The equation determining the lowest $R$ eigenstate is
\begin{equation}
    \left(\frac{K_f}{\gamma}-R+\frac{i}{2} D_f\right) \psi_0(x,u) = \left(\frac{K_f}{\gamma}-r_0+\frac{i}{2} D_f\right) \psi_0(x,u)=0,
\end{equation}
which can be written in terms of the charges for the decoupled theory \eqref{Eq:h0d0k0} (usual conformal mechanics) using \eqref{Eq:H1-D1-K1} and \eqref{Eq:Alg2} as 
\begin{equation}
    \left(a(u) H_0 + b(u) D_0  + c(u) K_0  -r_0 \right) \psi_0(x,u) = 0,
\end{equation}
where
\begin{align}\label{Eq:abc}
    a(u) & = \frac{f(u) (f(u)-i \gamma)}{\gamma f'(u)},\\
    b(u) &=\frac{1}{2}\left(i + \frac{f(u)}{\gamma}\left(-2 +\frac{(-i \gamma +f(u))f''(u)}{f'(u)^2}\right) \right),\\
    c(u) & = \frac{(2f'(u)^2 +(i\gamma -f(u))f''(u))(2 f'(u)^2-f(u)f''(u))}{4 \gamma f'(u)^3}.
\end{align}
Explicitly in the position basis we get the equation 
\begin{multline}
    \frac{a(u)}{2}\left(-\partial_x^2\psi_0(x,u)+\frac{q}{x^2} \psi_0(x,u) \right) + \frac{b(u)}{2} \left(-2 i x \partial_x\psi_0(x,u) - i \psi_0(x) \right) + \frac{c(u)}{2} x^2 \psi_0(x,u) \\= r_0 \psi_0(x,u),
\end{multline}
which can be readily solved  using the explicit expressions for $a,b,c$ from \eqref{Eq:abc} to obtain the solution \eqref{Eq:psi0} after normalization.

We note that
\begin{equation}
    \psi_1(x,u) =\frac{1}{\sqrt{2r_0}} L_+^R \psi_0(x,u)=\frac{1}{\sqrt{2r_0}}\left(\frac{K_f}{\gamma}-r_0-\frac{i}{2} D_f\right) \psi_0(x,u).
\end{equation}
Explicitly in the position basis this takes the following form
\begin{multline}
    \psi_1(x,u) =\frac{1}{\sqrt{2r_0}}\Bigg( \frac{\bar{a}(u)}{2}\left(-\partial_x^2\psi_0(x,u)+\frac{q}{x^2} \psi_0(x,u) \right) +\\ +\frac{\bar{b}(u)}{2} \left(-2 i x \partial_x\psi_0(x,u) - i \psi_0(x) \right) +\\ +\frac{\bar{c}(u)}{2} x^2 \psi_0(x,u) -r_0 \psi_0(x,u)\Bigg),
\end{multline}
where the bars denote complex conjugation. The above readily gives \eqref{Eq:psi1}. One can similarly obtain the higher eigenstates of $R$.

\section{Finding the time dependent phases}
\label{App:phase}
The equation for the time dependent phase ensuring that the $R$ eigenstates satisfy the time-dependent Schrodinger equation is
\begin{equation}\label{Eq:phase-2}
     \xi_n'(u) = \braket{\psi_n(u)| i \partial_u - \mathbb{H}|\psi_n(u) }
\end{equation}
This can be re-written as
\begin{equation}
    \xi_n'(u) = \frac{1}{c_+(r_0,r_0+n-1)^2}\braket{\psi_{n-1}(u)| L_{-}^R( i \partial_u - \mathbb{H}) L_+^R|\psi_{n-1}(u) }.
\end{equation}
The operators $L_{\pm}^R$ are linear combinations of the time invariant operators $H_f$, $K_f$ and $D_f$ with time-independent coefficients and therefore are themselves invariants of motion. This implies that
\begin{equation}
    i \partial_u L_+^R  - [\mathbb{H} , L_+^R] = 0.
\end{equation}
We can then simplify the expression for the phase as
\begin{align}
     \xi_n'(u) &= \frac{1}{c_+(r_0,r_0+n-1)^2}\braket{\psi_{n-1}(x,u)| L_{-}^R L_+^R( i \partial_u - \mathbb{H})|\psi_{n-1}(x,u) },\nonumber\\
               &= \braket{\psi_{n-1}(x,u)| ( i \partial_u - \mathbb{H})|\psi_{n-1}(x,u) } = \xi_{n-1}'(u),
\end{align}
where in the last line we have used the Hermitian adjoint of the expression $$L_{-}^R L_+^R \ket{\psi_{n-1}(x,u)} =c_+(r_0,r_0+n-1)^2 \ket{\psi_{n-1}(x,u)}$$(note that $c_-(r_0,r_0+n) = c_+(r_0,r_0+n-1)$). Repeating this procedure $n$ times we obtain
\begin{equation}
    \xi_n'(u)= \xi_{0}'(u) :=\xi'(u),
\end{equation}
completing the proof for the $n$ independence of the phases. To obtain $\xi(u)$ as given by \eqref{Eq:xi} explicitly, we need to use that
\begin{equation}
   \xi'(u)= \braket{\psi_0(u)| i \partial_u - \mathbb{H}|\psi_0(u) } = \frac{-2 r_0 \gamma f'(u)}{\gamma^2+f(u)^2},
\end{equation}
which can readily be integrated to obtain the phase  \eqref{Eq:xi}.

 \bibliographystyle{JHEP} 
 \bibliography{NAdS2LN}

\providecommand{\href}[2]{#2}\begingroup\raggedright\begin{thebibliography}{10}

\bibitem{Teitelboim:1983ux}
C.~Teitelboim, \emph{{Gravitation and Hamiltonian Structure in Two Space-Time
  Dimensions}}, \href{https://doi.org/10.1016/0370-2693(83)90012-6}{\emph{Phys.
  Lett. B} {\bfseries 126} (1983) 41}.

\bibitem{Jackiw:1984je}
R.~Jackiw, \emph{{Lower Dimensional Gravity}},
  \href{https://doi.org/10.1016/0550-3213(85)90448-1}{\emph{Nucl. Phys. B}
  {\bfseries 252} (1985) 343}.

\bibitem{Almheiri:2014cka}
A.~Almheiri and J.~Polchinski, \emph{{Models of AdS$_{2}$ backreaction and
  holography}}, \href{https://doi.org/10.1007/JHEP11(2015)014}{\emph{JHEP}
  {\bfseries 11} (2015) 014} [\href{https://arxiv.org/abs/1402.6334}{{\ttfamily
  1402.6334}}].

\bibitem{Maldacena:2016hyu}
J.~Maldacena and D.~Stanford, \emph{{Remarks on the Sachdev-Ye-Kitaev model}},
  \href{https://doi.org/10.1103/PhysRevD.94.106002}{\emph{Phys. Rev. D}
  {\bfseries 94} (2016) 106002}
  [\href{https://arxiv.org/abs/1604.07818}{{\ttfamily 1604.07818}}].

\bibitem{Jensen:2016pah}
K.~Jensen, \emph{{Chaos in AdS$_2$ Holography}},
  \href{https://doi.org/10.1103/PhysRevLett.117.111601}{\emph{Phys. Rev. Lett.}
  {\bfseries 117} (2016) 111601}
  [\href{https://arxiv.org/abs/1605.06098}{{\ttfamily 1605.06098}}].

\bibitem{Maldacena:2016upp}
J.~Maldacena, D.~Stanford and Z.~Yang, \emph{{Conformal symmetry and its
  breaking in two dimensional Nearly Anti-de-Sitter space}},
  \href{https://doi.org/10.1093/ptep/ptw124}{\emph{PTEP} {\bfseries 2016}
  (2016) 12C104} [\href{https://arxiv.org/abs/1606.01857}{{\ttfamily
  1606.01857}}].

\bibitem{Engelsoy:2016xyb}
J.~Engels\"oy, T.G.~Mertens and H.~Verlinde, \emph{{An investigation of
  AdS$_{2}$ backreaction and holography}},
  \href{https://doi.org/10.1007/JHEP07(2016)139}{\emph{JHEP} {\bfseries 07}
  (2016) 139} [\href{https://arxiv.org/abs/1606.03438}{{\ttfamily
  1606.03438}}].

\bibitem{Joshi:2019wgi}
L.K.~Joshi, A.~Mukhopadhyay and A.~Soloviev, \emph{{Time-dependent $NAdS_2$
  holography with applications}},
  \href{https://doi.org/10.1103/PhysRevD.101.066001}{\emph{Phys. Rev. D}
  {\bfseries 101} (2020) 066001}
  [\href{https://arxiv.org/abs/1901.08877}{{\ttfamily 1901.08877}}].

\bibitem{Mertens:2022irh}
T.G.~Mertens and G.J.~Turiaci, \emph{{Solvable models of quantum black holes: a
  review on Jackiw\textendash{}Teitelboim gravity}},
  \href{https://doi.org/10.1007/s41114-023-00046-1}{\emph{Living Rev. Rel.}
  {\bfseries 26} (2023) 4} [\href{https://arxiv.org/abs/2210.10846}{{\ttfamily
  2210.10846}}].

\bibitem{AEMM}
A.~Almheiri, N.~Engelhardt, D.~Marolf and H.~Maxfield, \emph{{The entropy of
  bulk quantum fields and the entanglement wedge of an evaporating black
  hole}}, \href{https://doi.org/10.1007/JHEP12(2019)063}{\emph{JHEP} {\bfseries
  12} (2019) 063} [\href{https://arxiv.org/abs/1905.08762}{{\ttfamily
  1905.08762}}].

\bibitem{AlmheiriQES}
A.~Almheiri, R.~Mahajan, J.~Maldacena and Y.~Zhao, \emph{{The Page curve of
  Hawking radiation from semiclassical geometry}},
  \href{https://doi.org/10.1007/JHEP03(2020)149}{\emph{JHEP} {\bfseries 03}
  (2020) 149} [\href{https://arxiv.org/abs/1908.10996}{{\ttfamily
  1908.10996}}].

\bibitem{Penington:2023dql}
G.~Penington and E.~Witten, \emph{{Algebras and States in JT Gravity}},
  \href{https://arxiv.org/abs/2301.07257}{{\ttfamily 2301.07257}}.

\bibitem{Kolchmeyer:2023gwa}
D.K.~Kolchmeyer, \emph{{von Neumann algebras in JT gravity}},
  \href{https://doi.org/10.1007/JHEP06(2023)067}{\emph{JHEP} {\bfseries 06}
  (2023) 067} [\href{https://arxiv.org/abs/2303.04701}{{\ttfamily
  2303.04701}}].

\bibitem{Jacobs2006}
K.~Jacobs and D.A.~Steck, \emph{A straightforward introduction to continuous
  quantum measurement},
  \href{https://doi.org/10.1080/00107510601101934}{\emph{Contemporary Physics}
  {\bfseries 47} (2006) 279}
  [\href{https://arxiv.org/abs/https://doi.org/10.1080/00107510601101934}{{\ttfamily
  https://doi.org/10.1080/00107510601101934}}].

\bibitem{Quanta12}
B.~Svensson, \emph{Pedagogical review of quantum measurement theory with an
  emphasis on weak measurements},
  \href{https://doi.org/10.12743/quanta.v2i1.12}{\emph{Quanta} {\bfseries 2}
  (2013) 18}.

\bibitem{Case:1950an}
K.M.~Case, \emph{{Singular potentials}},
  \href{https://doi.org/10.1103/PhysRev.80.797}{\emph{Phys. Rev.} {\bfseries
  80} (1950) 797}.

\bibitem{deAlfaro:1976vlx}
V.~de~Alfaro, S.~Fubini and G.~Furlan, \emph{{Conformal Invariance in Quantum
  Mechanics}}, \href{https://doi.org/10.1007/BF02785666}{\emph{Nuovo Cim. A}
  {\bfseries 34} (1976) 569}.

\bibitem{Claus:1998ts}
P.~Claus, M.~Derix, R.~Kallosh, J.~Kumar, P.K.~Townsend and A.~Van~Proeyen,
  \emph{{Black holes and superconformal mechanics}},
  \href{https://doi.org/10.1103/PhysRevLett.81.4553}{\emph{Phys. Rev. Lett.}
  {\bfseries 81} (1998) 4553}
  [\href{https://arxiv.org/abs/hep-th/9804177}{{\ttfamily hep-th/9804177}}].

\bibitem{Britto-Pacumio:1999dnb}
R.~Britto-Pacumio, J.~Michelson, A.~Strominger and A.~Volovich, \emph{{Lectures
  on Superconformal Quantum Mechanics and Multi-Black Hole Moduli Spaces}},
  \href{https://doi.org/10.1007/978-94-011-4303-5_6}{\emph{NATO Sci. Ser. C}
  {\bfseries 556} (2000) 255}
  [\href{https://arxiv.org/abs/hep-th/9911066}{{\ttfamily hep-th/9911066}}].

\bibitem{Nayak:2018qej}
P.~Nayak, A.~Shukla, R.M.~Soni, S.P.~Trivedi and V.~Vishal, \emph{{On the
  Dynamics of Near-Extremal Black Holes}},
  \href{https://doi.org/10.1007/JHEP09(2018)048}{\emph{JHEP} {\bfseries 09}
  (2018) 048} [\href{https://arxiv.org/abs/1802.09547}{{\ttfamily
  1802.09547}}].

\bibitem{Moitra:2018jqs}
U.~Moitra, S.P.~Trivedi and V.~Vishal, \emph{{Extremal and near-extremal black
  holes and near-CFT$_{1}$}},
  \href{https://doi.org/10.1007/JHEP07(2019)055}{\emph{JHEP} {\bfseries 07}
  (2019) 055} [\href{https://arxiv.org/abs/1808.08239}{{\ttfamily
  1808.08239}}].

\bibitem{Sachdev:2019bjn}
S.~Sachdev, \emph{{Universal low temperature theory of charged black holes with
  AdS$_2$ horizons}}, \href{https://doi.org/10.1063/1.5092726}{\emph{J. Math.
  Phys.} {\bfseries 60} (2019) 052303}
  [\href{https://arxiv.org/abs/1902.04078}{{\ttfamily 1902.04078}}].

\bibitem{Maldacena:1998uz}
J.M.~Maldacena, J.~Michelson and A.~Strominger, \emph{{Anti-de Sitter
  fragmentation}},
  \href{https://doi.org/10.1088/1126-6708/1999/02/011}{\emph{JHEP} {\bfseries
  02} (1999) 011} [\href{https://arxiv.org/abs/hep-th/9812073}{{\ttfamily
  hep-th/9812073}}].

\bibitem{PhysRevD.102.086008}
T.~Kibe, A.~Mukhopadhyay, H.~Swain and A.~Soloviev, \emph{$sl(2, r)$ lattices
  as information processors},
  \href{https://doi.org/10.1103/PhysRevD.102.086008}{\emph{Phys. Rev. D}
  {\bfseries 102} (2020) 086008}.

\bibitem{Kibe:2021gtw}
T.~Kibe, P.~Mandayam and A.~Mukhopadhyay, \emph{{Holographic spacetime, black
  holes and quantum error correcting codes: a review}},
  \href{https://doi.org/10.1140/epjc/s10052-022-10382-1}{\emph{Eur. Phys. J. C}
  {\bfseries 82} (2022) 463}
  [\href{https://arxiv.org/abs/2110.14669}{{\ttfamily 2110.14669}}].

\bibitem{Kibe:2023ixa}
T.~Kibe, S.~Mondkar, A.~Mukhopadhyay and H.~Swain, \emph{{Black hole
  complementarity from microstate models: a study of information replication
  and the encoding in the black hole interior}},
  \href{https://doi.org/10.1007/JHEP10(2023)096}{\emph{JHEP} {\bfseries 10}
  (2023) 096} [\href{https://arxiv.org/abs/2307.04799}{{\ttfamily
  2307.04799}}].

\bibitem{PhysRevD.48.3743}
L.~Susskind, L.~Thorlacius and J.~Uglum, \emph{The stretched horizon and black
  hole complementarity},
  \href{https://doi.org/10.1103/PhysRevD.48.3743}{\emph{Phys. Rev. D}
  {\bfseries 48} (1993) 3743}.

\bibitem{PhysRevD.50.2700}
L.~Susskind and J.~Uglum, \emph{Black hole entropy in canonical quantum gravity
  and superstring theory},
  \href{https://doi.org/10.1103/PhysRevD.50.2700}{\emph{Phys. Rev. D}
  {\bfseries 50} (1994) 2700}.

\bibitem{Harlow:2014yka}
D.~Harlow, \emph{{Jerusalem Lectures on Black Holes and Quantum Information}},
  \href{https://doi.org/10.1103/RevModPhys.88.015002}{\emph{Rev. Mod. Phys.}
  {\bfseries 88} (2016) 015002}
  [\href{https://arxiv.org/abs/1409.1231}{{\ttfamily 1409.1231}}].

\bibitem{Raju:2020smc}
S.~Raju, \emph{{Lessons from the Information Paradox}}, {\emph{arXiv}
  {\bfseries 2012.05770} (2020) }.

\bibitem{Ghosh:2019rcj}
A.~Ghosh, H.~Maxfield and G.J.~Turiaci, \emph{{A universal Schwarzian sector in
  two-dimensional conformal field theories}},
  \href{https://doi.org/10.1007/JHEP05(2020)104}{\emph{JHEP} {\bfseries 05}
  (2020) 104} [\href{https://arxiv.org/abs/1912.07654}{{\ttfamily
  1912.07654}}].

\bibitem{Iliesiu:2020qvm}
L.V.~Iliesiu and G.J.~Turiaci, \emph{{The statistical mechanics of
  near-extremal black holes}},
  \href{https://doi.org/10.1007/JHEP05(2021)145}{\emph{JHEP} {\bfseries 05}
  (2021) 145} [\href{https://arxiv.org/abs/2003.02860}{{\ttfamily
  2003.02860}}].

\bibitem{Heydeman:2020hhw}
M.~Heydeman, L.V.~Iliesiu, G.J.~Turiaci and W.~Zhao, \emph{{The statistical
  mechanics of near-BPS black holes}},
  \href{https://doi.org/10.1088/1751-8121/ac3be9}{\emph{J. Phys. A} {\bfseries
  55} (2022) 014004} [\href{https://arxiv.org/abs/2011.01953}{{\ttfamily
  2011.01953}}].

\bibitem{Sen:2008vm}
A.~Sen, \emph{{Quantum Entropy Function from AdS(2)/CFT(1) Correspondence}},
  \href{https://doi.org/10.1142/S0217751X09045893}{\emph{Int. J. Mod. Phys. A}
  {\bfseries 24} (2009) 4225}
  [\href{https://arxiv.org/abs/0809.3304}{{\ttfamily 0809.3304}}].

\bibitem{Gross:1973id}
D.J.~Gross and F.~Wilczek, \emph{{Ultraviolet Behavior of Nonabelian Gauge
  Theories}}, \href{https://doi.org/10.1103/PhysRevLett.30.1343}{\emph{Phys.
  Rev. Lett.} {\bfseries 30} (1973) 1343}.

\bibitem{Politzer:1973fx}
H.D.~Politzer, \emph{{Reliable Perturbative Results for Strong Interactions?}},
  \href{https://doi.org/10.1103/PhysRevLett.30.1346}{\emph{Phys. Rev. Lett.}
  {\bfseries 30} (1973) 1346}.

\bibitem{PhysRevLett.18.510}
H.R.~Lewis, \emph{Classical and quantum systems with time-dependent
  harmonic-oscillator-type hamiltonians},
  \href{https://doi.org/10.1103/PhysRevLett.18.510}{\emph{Phys. Rev. Lett.}
  {\bfseries 18} (1967) 510}.

\bibitem{choi2004coherent}
J.R.~Choi, \emph{Coherent states of general time-dependent harmonic
  oscillator}, {\emph{Pramana} {\bfseries 62} (2004) 13}.

\bibitem{Andrzejewski:2015jya}
K.~Andrzejewski, \emph{{Quantum conformal mechanics emerging from unitary
  representations of SL(2,$\mathbb{R}$)}},
  \href{https://doi.org/10.1016/j.aop.2016.01.020}{\emph{Annals Phys.}
  {\bfseries 367} (2016) 227}
  [\href{https://arxiv.org/abs/1506.05596}{{\ttfamily 1506.05596}}].

\bibitem{Michelson:1999zf}
J.~Michelson and A.~Strominger, \emph{{The Geometry of (super)conformal quantum
  mechanics}}, \href{https://doi.org/10.1007/PL00005528}{\emph{Commun. Math.
  Phys.} {\bfseries 213} (2000) 1}
  [\href{https://arxiv.org/abs/hep-th/9907191}{{\ttfamily hep-th/9907191}}].

\bibitem{Hayden_2007}
P.~Hayden and J.~Preskill, \emph{Black holes as mirrors: quantum information in
  random subsystems},
  \href{https://doi.org/10.1088/1126-6708/2007/09/120}{\emph{Journal of High
  Energy Physics} {\bfseries 2007} (2007) 120–120}.

\bibitem{Banerjee:2017ozx}
S.~Banerjee, N.~Gaddam and A.~Mukhopadhyay, \emph{{Illustrated study of the
  semiholographic nonperturbative framework}},
  \href{https://doi.org/10.1103/PhysRevD.95.066017}{\emph{Phys. Rev. D}
  {\bfseries 95} (2017) 066017}
  [\href{https://arxiv.org/abs/1701.01229}{{\ttfamily 1701.01229}}].

\bibitem{Kurkela:2018dku}
A.~Kurkela, A.~Mukhopadhyay, F.~Preis, A.~Rebhan and A.~Soloviev, \emph{{Hybrid
  Fluid Models from Mutual Effective Metric Couplings}},
  \href{https://doi.org/10.1007/JHEP08(2018)054}{\emph{JHEP} {\bfseries 08}
  (2018) 054} [\href{https://arxiv.org/abs/1805.05213}{{\ttfamily
  1805.05213}}].

\bibitem{Ecker:2018ucc}
C.~Ecker, A.~Mukhopadhyay, F.~Preis, A.~Rebhan and A.~Soloviev, \emph{{Time
  evolution of a toy semiholographic glasma}},
  \href{https://doi.org/10.1007/JHEP08(2018)074}{\emph{JHEP} {\bfseries 08}
  (2018) 074} [\href{https://arxiv.org/abs/1806.01850}{{\ttfamily
  1806.01850}}].

\bibitem{Mondkar:2021qsf}
S.~Mondkar, A.~Mukhopadhyay, A.~Rebhan and A.~Soloviev, \emph{{Quasinormal
  modes of a semi-holographic black brane and thermalization}},
  \href{https://doi.org/10.1007/JHEP11(2021)080}{\emph{JHEP} {\bfseries 11}
  (2021) 080} [\href{https://arxiv.org/abs/2108.02788}{{\ttfamily
  2108.02788}}].

\end{thebibliography}\endgroup

\end{document}